\input harvmac.tex



\def\unlockat{\catcode`\@=11}
\def\lockat{\catcode`\@=12}

\unlockat

\def\newsec#1{\global\advance\secno by1\message{(\the\secno. #1)}
\global\subsecno=0\global\subsubsecno=0\eqnres@t\noindent
{\bf\the\secno. #1}
\writetoca{{\secsym} {#1}}\par\nobreak\medskip\nobreak}
\global\newcount\subsecno \global\subsecno=0
\def\subsec#1{\global\advance\subsecno
by1\message{(\secsym\the\subsecno. #1)}
\ifnum\lastpenalty>9000\else\bigbreak\fi\global\subsubsecno=0
\noindent{\it\secsym\the\subsecno. #1}
\writetoca{\string\quad {\secsym\the\subsecno.} {#1}}
\par\nobreak\medskip\nobreak}
\global\newcount\subsubsecno \global\subsubsecno=0
\def\subsubsec#1{\global\advance\subsubsecno by1
\message{(\secsym\the\subsecno.\the\subsubsecno. #1)}
\ifnum\lastpenalty>9000\else\bigbreak\fi
\noindent\quad{\secsym\the\subsecno.\the\subsubsecno.}{#1}
\writetoca{\string\qquad{\secsym\the\subsecno.\the\subsubsecno.}{#1}}
\par\nobreak\medskip\nobreak}

\def\subsubseclab#1{\DefWarn#1\xdef
#1{\noexpand\hyperref{}{subsubsection}%
{\secsym\the\subsecno.\the\subsubsecno}%
{\secsym\the\subsecno.\the\subsubsecno}}%
\writedef{#1\leftbracket#1}\wrlabeL{#1=#1}}
\lockat

\def\IL{\relax{\rm I\kern-.18em L}}
\def\IH{\relax{\rm I\kern-.18em H}}
\def\IR{\relax{\rm I\kern-.18em R}}
\def\IC{\relax\hbox{$\inbar\kern-.3em{\rm C}$}}
\def\IZ{\relax\ifmmode\mathchoice
{\hbox{\cmss Z\kern-.4em Z}}{\hbox{\cmss Z\kern-.4em Z}}
{\lower.9pt\hbox{\cmsss Z\kern-.4em Z}}
{\lower1.2pt\hbox{\cmsss Z\kern-.4em Z}}\else{\cmss Z\kern-.4em
Z}\fi}

\def\CN {{\cal N}}
\def\CR {{\cal R}}

\def\CF {{\cal F}}

\def\CZ {{\cal Z}}
\def\CE {{\cal E}}
\def\CG {{\cal G}}

\def\CC {{\cal C}}

\def\CS {{\cal S}}
\def\CA{{\cal A}}


\def\CN {{\cal N}}

\def\CE{{\cal E }}

\def\CZ {{\cal Z }}
\def\CS {{\cal S }}
\def\ch{{\rm ch}}

\def\Det{{\rm Det}}

\def\ib{{\bar i}}
\def\kb{{\bar k}}
\def\IZ{\relax\ifmmode\mathchoice
{\hbox{\cmss Z\kern-.4em Z}}{\hbox{\cmss Z\kern-.4em Z}}
{\lower.9pt\hbox{\cmsss Z\kern-.4em Z}}
{\lower1.2pt\hbox{\cmsss Z\kern-.4em Z}}\else{\cmss Z\kern-.4em
Z}\fi}

\def\zb {\bar{z}}

\font\manual=manfnt \def\dbend{\lower3.5pt\hbox{\manual\char127}}

\font\cmss=cmss10 \font\cmsss=cmss10 at 7pt
\def\IZ{\relax\ifmmode\mathchoice
{\hbox{\cmss Z\kern-.4em Z}}{\hbox{\cmss Z\kern-.4em Z}}
{\lower.9pt\hbox{\cmsss Z\kern-.4em Z}}
{\lower1.2pt\hbox{\cmsss Z\kern-.4em Z}}\else{\cmss Z\kern-.4em
Z}\fi}
\def\half {{1\over 2}}

\def\p{\partial}
\def\pb{\bar{\partial}}

\def\CN {{\cal N}}

\def\CE{{\cal E }}

\def\CZ {{\cal Z }}
\def\CS {{\cal S }}
\def\ch{{\rm ch}}
\def\Det{{\rm Det}}

\def\ib{{\bar i}}


\def\IZ{\relax\ifmmode\mathchoice
{\hbox{\cmss Z\kern-.4em Z}}{\hbox{\cmss Z\kern-.4em Z}}
{\lower.9pt\hbox{\cmsss Z\kern-.4em Z}}
{\lower1.2pt\hbox{\cmsss Z\kern-.4em Z}}\else{\cmss Z\kern-.4em
Z}\fi}
\def\IB{\relax{\rm I\kern-.18em B}}
\def\IC{{\relax\hbox{$\inbar\kern-.3em{\rm C}$}}}
\def\ID{\relax{\rm I\kern-.18em D}}
\def\IE{\relax{\rm I\kern-.18em E}}
\def\IF{\relax{\rm I\kern-.18em F}}
\def\IG{\relax\hbox{$\inbar\kern-.3em{\rm G}$}}
\def\IGa{\relax\hbox{${\rm I}\kern-.18em\Gamma$}}
\def\IH{\relax{\rm I\kern-.18em H}}
\def\II{\relax{\rm I\kern-.18em I}}
\def\IK{\relax{\rm I\kern-.18em K}}
\def\IP{\relax{\rm I\kern-.18em P}}

\def\jb{{\bar j}}
\def\k{K\"ahler\ }
\def\lieg{{\underline{\bf g}}}

\def\inbar{\,\vrule height1.5ex width.4pt depth0pt}
\def\p{\partial}
\def\pb{{\bar \p}}

\font\cmss=cmss10 \font\cmsss=cmss10 at 7pt
\def\IR{\relax{\rm I\kern-.18em R}}
\def\pbar{\bar{\p}}

\def\Td{{\rm Td}}
\def\Tr{{\rm Tr}}

\def\wzwt{$WZW_2$}
\def\wzwf{$WZW_4$}

\def\zb {{\bar{z}}}

\def\boxit#1{\vbox{\hrule\hbox{\vrule\kern8pt
\vbox{\hbox{\kern8pt}\hbox{\vbox{#1}}\hbox{\kern8pt}}
\kern8pt\vrule}\hrule}}
\def\mathboxit#1{\vbox{\hrule\hbox{\vrule\kern8pt\vbox{\kern8pt
\hbox{$\displaystyle #1$}\kern8pt}\kern8pt\vrule}\hrule}}


\def\bqa{\bar Q^{\dot A} }

\def\jb{{\bar j}}

\def\lieg{{\underline{\bf g}}}

\def\inbar{\,\vrule height1.5ex width.4pt depth0pt}

\def\p{\partial}
\def\pab{\pb_{\bar a} }
\def\pb{{\bar \p}}

\font\cmss=cmss10 \font\cmsss=cmss10 at 7pt
\def\IR{\relax{\rm I\kern-.18em R}}
\def\pbar{\bar{\p}}

\def\zb {{\bar{z}}}


\def\a1{{\cal A}^{1,1}}


\def\doz{\delta^{1,0} }
\def\dzo{\delta^{0,1} }
\def\dge{\delta}

%
\lref\cangemi{ D. Cangemi, ``Self-dual Yang-Mills
Theory and One-Loop Like-Helicity QCD
Multi-Gluon Amplitudes,'' hep-th/9605208. }
\lref\zumino{B. Zumino,
``Chiral anomalies and differential geometry,''
in {\it Relativity, Groups and Topology II},
proceedings of the
Les Houches summer school, Bryce S DeWitt and Raymond Stora, eds.
North-Holland, 1984.}
\lref\anomrev{For reviews see {\it Symposium on Anomalies, Geometry and
Topology }
William A. Bardeen and Alan R.
White, eds. World Scientific, 1985, and
L. Alvarez-Gaum\'e and P. Ginsparg,
``The structure of gauge and gravitational anomalies,''
Ann. Phys. {\bf 161} (1985) 423;
L. Alvarez-Gaum\'e,
``An introduction to anomalies,''
Lectures given at Int.
School on Mathematical Physics, Erice, Italy, Jul 1-14,
1985.
Published in Erice School Math.Phys.1985:93
}
\lref\agwitt{L. Alvarez-Gaum\'e and E. Witten,
``Gravitational Anomalies,'' Nucl. Phys.
{\bf B234}(1983) 269}
\lref\blzh{A. Belavin, V. Zakharov, ``Yang-Mills Equations as inverse
scattering
problem''Phys. Lett. B73, (1978) 53}
\lref\afs{Alekseev, Faddeev, Shatashvili,  }
\lref\bost{L. Alvarez-Gaume, J.B. Bost , G. Moore, P. Nelson, C.
Vafa,
``Bosonization on higher genus Riemann surfaces,''
Commun.Math.Phys.112:503,1987}
\lref\agmv{L. Alvarez-Gaum\'e,
C. Gomez, G. Moore,
and C. Vafa, ``Strings in the Operator Formalism,''
Nucl. Phys. {\bf 303}(1988)455}
\lref\atiyah{M. Atiyah, ``Green's Functions for
Self-Dual Four-Manifolds,'' Adv. Math. Suppl.
{\bf 7A} (1981)129}

\lref\AHS{M.~ Atiyah, N.~ Hitchin and I.~ Singer, ``Self-Duality in
Four-Dimensional
Riemannian Geometry", Proc. Royal Soc. (London) {\bf A362} (1978)
425-461.}
\lref\fmlies{M. F. Atiyah and I. M. Singer,
``The index of elliptic operators IV,'' Ann. Math. {\bf 93}(1968)119}
\lref\bagger{E. Witten and J. Bagger, Phys. Lett.
{\bf 115B}(1982)202}
\lref\banks{T. Banks, ``Vertex Operators in 2D Dimensions,''
hep-th/9503145   }
\lref\bardeen{W.~ A.~ Bardeen, ``Self-Dual Yang-Mills, Integrability
and Multi-Parton Amplitudes'',
Fermilab - Conf - -95-379-T, Aug 1995,
Presented at Yukawa International Seminar '95:
`From the Standard Model to Grand
Unified Theories', Kyoto, Japan, 21-25 Aug 1995. }
\lref\berk{N. Berkovits, ``Super-Poincare Invariant Superstring Field Theory''
hep-th/9503099 }
\lref\biquard{O. Biquard, ``Sur les fibr\'es paraboliques
sur une surface complexe,'' to appear in J. Lond. Math.
Soc.}
\lref\bjsv{hep-th/9501096,
Topological Reduction of 4D SYM to 2D $\sigma$--Models,
 M. Bershadsky, A. Johansen, V. Sadov and C. Vafa }
\lref\BlThlgt{M.~ Blau and G.~ Thompson, ``Lectures on 2d Gauge
Theories: Topological Aspects and Path
Integral Techniques", Presented at the
Summer School in Hogh Energy Physics and
Cosmology, Trieste, Italy, 14 Jun - 30 Jul
1993, hep-th/9310144.}
\lref\bpz{A.A. Belavin, A.M. Polyakov, A.B. Zamolodchikov,
``Infinite conformal symmetry in two-dimensional quantum
field theory,'' Nucl.Phys.B241:333,1984}
\lref\BGV{N. Berline, E. Getzler, and M. Vergne,
{\it Heat Kernels and Dirac Operators} Springer}
\lref\braam{P.J. Braam, A. Maciocia, and A. Todorov,
``Instanton moduli as a novel map from tori to
K3-surfaces,'' Inven. Math. {\bf 108} (1992) 419}
\lref\cllnhrvy{Callan and Harvey, Nucl
Phys. {\bf B250}(1985)427}
\lref\CMR{ For a review, see
S. Cordes, G. Moore, and S. Ramgoolam,
`` Lectures on 2D Yang Mills theory, Equivariant
Cohomology, and Topological String Theory,''
Lectures presented at the 1994 Les Houches Summer School
 ``Fluctuating Geometries in Statistical Mechanics and Field
Theory.''
and at the Trieste 1994 Spring school on superstrings.
hep-th/9411210, or see http://xxx.lanl.gov/lh94}
\lref\devchand{Ch. Devchand and V. Ogievetsky,
``Four dimensional integrable theories,'' hep-th/9410147}
\lref\devchandi{
Ch. Devchand and A.N. Leznov,
``B \"acklund transformation for supersymmetric self-dual theories for
semisimple
gauge groups and a hierarchy of $A_1$ solutions,'' hep-th/9301098,
Commun. Math. Phys. {\bf 160} (1994) 551}
\lref\dnld{S. Donaldson, ``Anti self-dual Yang-Mills
connections over complex  algebraic surfaces and stable
vector bundles,'' Proc. Lond. Math. Soc,
{\bf 50} (1985)1}

\lref\DoKro{S.K.~ Donaldson and P.B.~ Kronheimer,
{\it The Geometry of Four-Manifolds},
Clarendon Press, Oxford, 1990.}
\lref\donii{
S. Donaldson, Duke Math. J. , {\bf 54} (1987) 231. }

\lref\elitzur{S. Elitzur, G. Moore,
A. Schwimmer, and N. Seiberg,
``Remarks on the Canonical Quantization of the Chern-Simons-
Witten Theory,'' Nucl. Phys. {\bf B326}(1989)108;
G. Moore and N. Seiberg,
``Lectures on Rational Conformal Field Theory,''
, in {\it Strings'89},Proceedings
of the Trieste Spring School on Superstrings,
3-14 April 1989, M. Green, et. al. Eds. World
Scientific, 1990}
\lref\etingof{P.I. Etingof and I.B. Frenkel,
``Central Extensions of Current Groups in
Two Dimensions,'' Commun. Math.
Phys. {\bf 165}(1994) 429}

\lref\evans{M. Evans, F. G\"ursey, V. Ogievetsky,
``From 2D conformal to 4D self-dual theories:
Quaternionic analyticity,''
Phys. Rev. {\bf D47}(1993)3496}
\lref\fs{L. Faddeev and S. Shatashvili, Theor. Math. Fiz., 60 (1984)206\semi
L. Faddeev, Phys. Lett. B145 (1984) 81.\semi
J. Mickelsson, Commun. Math. Phys. , 97 (1985) 361.}
\lref\fsi{ L. Faddeev, Phys. Lett. B145 (1984) 81.}
\lref\fz{I. Frenkel, I. Singer, unpublished.}
\lref\fk{I. Frenkel and B. Khesin, ``Four dimensional
realization of two dimensional current groups,'' Yale
preprint, July 1995, to appear in Commun. Math. Phys.}
\lref\FMS{D. Friedan, E. Martinec,  and S. Shenker,
``Conformal Invariance, Supersymmetry, and String Theory,''
Nucl.Phys. {\bf B271} (1986) 93.}
\lref\galperin{A. Galperin, E. Ivanov, V. Ogievetsky,
E. Sokatchev, Ann. Phys. {\bf 185}(1988) 1}
\lref\gwdzki{K. Gawedzki, ``Topological Actions in Two-Dimensional
Quantum Field Theories,'' in {\it Nonperturbative
Quantum Field Theory}, G. 't Hooft, A. Jaffe, et. al. , eds. ,
Plenum 1988}
\lref\gmps{A. Gerasimov, A. Morozov, M. Olshanetskii,
 A. Marshakov, S. Shatashvili ,``
Wess-Zumino-Witten model as a theory of
free fields,'' Int. J. Mod. Phys. A5 (1990) 2495-2589
 }
\lref\gerasimov{A. Gerasimov, ``Localization in
GWZW and Verlinde formula,'' hepth/9305090}
\lref\ginzburg{V. Ginzburg, M. Kapranov, and E. Vasserot,
``Langlands Dualtiy for Surfaces,'' IAS preprint}
\lref\giveon{hep-th/9502057,
 S-Duality in N=4 Yang-Mills Theories with General Gauge Groups,
 Luciano Girardello, Amit Giveon, Massimo Porrati, and Alberto
Zaffaroni
}

\lref\gottsh{L. Gottsche, Math. Ann. 286 (1990)193}
\lref\gothuy{L. G\"ottsche and D. Huybrechts,
``Hodge numbers of moduli spaces of stable
bundles on $K3$ surfaces,'' alg-geom/9408001}
\lref\GrHa{P.~ Griffiths and J.~ Harris, {\it Principles of
Algebraic
geometry},
p. 445, J.Wiley and Sons, 1978. }
\lref\ripoff{I. Grojnowski, ``Instantons and
affine algebras I: the Hilbert scheme and
vertex operators,'' alg-geom/9506020.}
\lref\adhmfk{I. Grojnowski,
A. Losev, G. Moore, N. Nekrasov, S. Shatashvili,
``ADHM and the Frenkel-Kac construction,'' in preparation}

\lref\hitchin{N. Hitchin, ``Polygons and gravitons,''
Math. Proc. Camb. Phil. Soc, (1979){\bf 85} 465}

\lref\hklr{Hitchin, Karlhede, Lindstrom, and Rocek,
``Hyperkahler metrics and supersymmetry,''
Commun. Math. Phys. {\bf 108}(1987)535}
\lref\hirz{F. Hirzebruch and T. Hofer, Math. Ann. 286 (1990)255}
\lref\hms{hep-th/9501022,
 Reducing $S$- duality to $T$- duality, J. A. Harvey, G. Moore and A.
Strominger}
\lref\johansen{A. Johansen,
``Twisting of $N=1$ susy gauge theories and
heterotic topological theories,''
 Int.J.Mod.Phys.A10:4325-4358,1995, hep-th/9403017\semi
``Infinite Conformal
Algebras in Supersymmetric Theories on
Four Manifolds,'' Nucl.Phys.B436:291-341,1995,
hep-th/9407109\semi ``Realization of $W_{1+\infty}$
and Virasoro algebras in supersymmetric theories on
four manifolds,''
 Mod.Phys.Lett.A9:2611-2622,1994, hep-th/9406156}
\lref\kronheimer{P. Kronheimer, ``The construction of ALE spaces as
hyper-kahler quotients,'' J. Diff. Geom. {\bf 28}1989)665}
\lref\kricm{P. Kronheimer, ``Embedded surfaces in
4-manifolds,'' Proc. Int. Cong. of
Math. (Kyoto 1990) ed. I. Satake, Tokyo, 1991}

\lref\KN{Kronheimer and Nakajima,  ``Yang-Mills instantons
on ALE gravitational instantons,''  Math. Ann.
{\bf 288}(1990)263}
\lref\krmw{P. Kronheimer and T. Mrowka,
``Gauge theories for embedded surfaces I,''
Topology {\bf 32} (1993) 773,
``Gauge theories for embedded surfaces II,''
preprint.}

\lref\hypvol{A. Losev, G. Moore, N. Nekrasov, S. Shatashvili,
``Localization for Hyperkahler Quotients,
Integration over Instanton Moduli,
and ALE Spaces,'' in preparation}
\lref\fdrcft{A. Losev, G. Moore, N. Nekrasov, S. Shatashvili, in
preparation.}
\lref\avatar{A. Losev, G. Moore, N. Nekrasov, S. Shatashvili,
``Four-Dimensional Avatars of 2D RCFT,''
hep-th/9509151.}
\lref\cocycle{A. Losev, G. Moore, N. Nekrasov, S. Shatashvili,
``Central Extensions of Gauge Groups Revisited,''
hep-th/9511185.}
\lref\kutmart{D.~ Kutasov,  E.~ Martinec, ``New Principles for String/Membrane
Unification.''
hep-th/9602049\semi
D.~ Kutasov  ,  E.~ Martinec  ,  M.~ O'~Loughlin,
``Vacua of M-theory and N=2 strings,'' hep-th/9603116}

\lref\maciocia{A. Maciocia, ``Metrics on the moduli
spaces of instantons over Euclidean 4-Space,''
Commun. Math. Phys. {\bf 135}(1991) , 467}
\lref\marcus{N. Marcus, ``A tour through
$N=2$ strings,'' hep-th/9211059}
\lref\mickold{J. Mickelsson, Commun. Math. Phys. , 97 (1985) 361.}
\lref\mick{J. Mickelsson, ``Kac-Moody groups,
topology of the Dirac determinant bundle and
fermionization,'' Commun. Math. Phys., {\bf 110} (1987) 173.}

\lref\milnor{J. Milnor, ``A unique decomposition
theorem for 3-manifolds,'' Amer. Jour. Math, (1961) 1}
\lref\taming{G. Moore and N. Seiberg,
``Taming the conformal zoo,'' Phys. Lett.
{\bf 220 B} (1989) 422}
\lref\nair{V.P.Nair, ``K\"ahler-Chern-Simons Theory'', hep-th/9110042
}
\lref\ns{V.P. Nair and Jeremy Schiff,
``Kahler Chern Simons theory and symmetries of
antiselfdual equations'' Nucl.Phys.B371:329-352,1992;
``A Kahler Chern-Simons theory and quantization of the
moduli of antiselfdual instantons,''
Phys.Lett.B246:423-429,1990,
``Topological gauge theory and twistors,''
Phys.Lett.B233:343,1989}
\lref\nakajima{H. Nakajima, ``Homology of moduli
spaces of instantons on ALE Spaces. I'' J. Diff. Geom.
{\bf 40}(1990) 105; ``Instantons on ALE spaces,
quiver varieties, and Kac-Moody algebras,'' preprint,
``Gauge theory on resolutions of simple singularities
and affine Lie algebras,'' preprint.}
\lref\nakheis{H.Nakajima, ``Heisenberg algebra and Hilbert schemes of points on
projective surfaces ,'' alg-geom/9507012}
\lref\ogvf{H. Ooguri and C. Vafa, ``Self-Duality
and $N=2$ String Magic,'' Mod.Phys.Lett. {\bf A5} (1990) 1389-1398\semi
``Geometry
of$N=2$ Strings,'' Nucl.Phys. {\bf B361}  (1991) 469-518\semi
``$N=2$ Heterotic Strings, Nucl.Phys.B367:83-104,1991 }
\lref\park{J.-S. Park, ``Holomorphic Yang-Mills theory on compact
Kahler
manifolds,'' hep-th/9305095; Nucl. Phys. {\bf B423} (1994) 559\semi
J.-S.~ Park, ``$N=2$ Topological Yang-Mills Theory on Compact
K\"ahler
Surfaces", Commun. Math, Phys. {\bf 163} (1994) 113\semi
J.-S.~ Park, ``$N=2$ Topological Yang-Mills Theories and Donaldson
Polynomials", hep-th/9404009}
\lref\parki{S. Hyun and J.-S. Park,
``Holomorphic Yang-Mills Theory and Variation
of the Donaldson Invariants,'' hep-th/9503036}
\lref\pohl{Pohlmeyer, Commun.
Math. Phys. {\bf 72}(1980)37}
\lref\phys{This interpretation has been
widely discussed in the physics literature.
For examples, see:
L. Alvarez-Gaum\'e and P. Ginsparg,
``The structure of gauge and gravitational anomalies,''
Ann. Phys. {\bf 161} (1985)423\semi G. Moore and
P. Nelson,
``The Aeteology of Sigma Model Anomalies,''
Comm. Math. Phys. {\bf 100}(1985)83.
}
\lref\pwf{A.~ M.~ Polyakov and P.~ B.~ Wiegmann,
Phys. Lett. {\bf B131}(1983)121}
\lref\plh{A.~ M.~ Polyakov, ``Two dimensional quantum
gravity: superconductivity at high $T_{c}$'', Les Houches lectures,
In *Les Houches 1988, Proceedings, Fields, strings and critical phenomena*
305-368}
\lref\prseg{Pressley and Segal, Loop Groups}
\lref\rade{J. Rade, ``Singular Yang-Mills fields. Local
theory I. '' J. reine ang. Math. , {\bf 452}(1994)111; {\it ibid}
{\bf 456}(1994)197; ``Singular Yang-Mills
fields-global theory,'' Intl. J. of Math. {\bf 5}(1994)491.}
\lref\segal{G. Segal, The definition of CFT}
\lref\seiberg{hep-th/9407087,
Monopole Condensation, And Confinement In $N=2$ Supersymmetric
Yang-Mills
Theory, N. Seiberg and E. Witten;
hep-th/9408013,  Nathan Seiberg;
hep-th/9408099,
Monopoles, Duality and Chiral Symmetry Breaking in
N=2 Supersymmetric QCD, N. Seiberg and E. Witten;
hep-th/9408155,
Phases of N=1 supersymmetric gauge theories in four dimensions, K. Intriligator
and N. Seiberg; hep-ph/9410203,
Proposal for a Simple Model of Dynamical SUSY Breaking, by K.
Intriligator, N.
Seiberg, and S. H. Shenker;
hep-th/9411149,
 Electric-Magnetic Duality in Supersymmetric Non-Abelian Gauge
Theories,
 N. Seiberg; hep-th/9503179 Duality, Monopoles, Dyons, Confinement and Oblique
Confinement in Supersymmetric $SO(N_c)$ Gauge Theories,
K. Intriligator and N. Seiberg}
\lref\sen{A. Sen,
hep-th/9402032, Dyon-Monopole bound states, selfdual harmonic
forms on the multimonopole moduli space and $SL(2,Z)$
invariance,'' }
\lref\shatashi{S. Shatashvili,
Theor. and Math. Physics, 71, 1987, p. 366}
\lref\thooft{G. 't Hooft , ``A property of electric and
magnetic flux in nonabelian gauge theories,''
Nucl.Phys.B153:141,1979}
\lref\vafa{C. Vafa, ``Conformal theories and punctured
surfaces,'' Phys.Lett.199B:195,1987 }
\lref\VaWi{C.~ Vafa and E.~ Witten, ``A Strong Coupling Test of
$S$-Duality",
hep-th/9408074.}
\lref\vrlsq{E. Verlinde and H. Verlinde,
``Conformal Field Theory and Geometric Quantization,''
in {\it Strings'89},Proceedings
of the Trieste Spring School on Superstrings,
3-14 April 1989, M. Green, et. al. Eds. World
Scientific, 1990}
\lref\mwxllvrld{E. Verlinde, ``Global Aspects of
Electric-Magnetic Duality,'' hep-th/9506011}
\lref\wrdhd{R. Ward, Nucl. Phys. {\bf B236}(1984)381}
\lref\ward{Ward and Wells, {\it Twistor Geometry and
Field Theory}, CUP }
\lref\wittenwzw{E. Witten, ``Nonabelian bosonization in
two dimensions,'' Commun. Math. Phys. {\bf 92} (1984)455 }
\lref\grssmm{E. Witten, ``Quantum field theory,
grassmannians and algebraic curves,'' Commun.Math.Phys.113:529,1988}
\lref\wittjones{E. Witten, ``Quantum field theory and the Jones
polynomial,'' Commun.  Math. Phys.}
\lref\wittentft{E.~ Witten, ``Topological Quantum Field Theory",
Commun. Math. Phys. {\bf 117} (1988) 353.}
\lref\Witdgt{ E.~ Witten, ``On Quantum gauge theories in two
dimensions,''
Commun. Math. Phys. {\bf  141}  (1991) 153.}
\lref\Witfeb{E.~ Witten, ``Supersymmetric Yang-Mills Theory On A
Four-Manifold,'' J. Math. Phys. {\bf 35} (1994) 5101.}
\lref\Witr{E.~ Witten, ``Introduction to Cohomological Field
Theories",
Lectures at Workshop on Topological Methods in Physics, Trieste,
Italy,
Jun 11-25, 1990, Int. J. Mod. Phys. {\bf A6} (1991) 2775.}
\lref\wittabl{E. Witten,  ``On S-Duality in Abelian Gauge Theory,''
hep-th/9505186}
\lref\doniii{S. Donaldson, ``Anti-Self-Dual Yang-Mills
Connections over Complex Algebraic Surfaces and
Stable Vector Bundles,'' Prod. Lond. Math. Soc.
{\bf 50}(1985)1. }
\lref\donii{
S. Donaldson, ``Infinite Determinants, Stable
Bundles, and Curvature,''
Duke Math. J. , {\bf 54} (1987) 231. }
\lref\BGS{J. -M. Bismut, H. Gillet, and C. Soul\'e,
``Analytic Torsion and Holomorphic Determinant
Bundles, I.II.III''
CMP {\bf 115}(1988)49-78;79-126;301-351}
\lref\luis{L.~ Alvarez-Gaum\'e, ``Supersymmetry and Atyah-Singer index
theorem'', Commun.Math.Phys.90:161,1983}
\lref\luisp{L.~ Alvarez-Gaum\'e,
``An Introduction to Anomalies,''
Lectures given at Int. School on Mathematical Physics, Erice, Italy, Jul 1-14,
1985.
Published in Erice School Math.Phys.1985:93}
\lref\fw{D. Friedan and P. Windey,
``Supersymmetric derivation of the Atyah-Singer index
and the chiral anomaly,''  Nucl.Phys.B235:395,1984}
\lref\knizhnik{V.G. Knizhnik,
``Analytic fields on Riemann surfaces,''
In *Nishinomiya 1987, Proceedings, Quantum string theory* 120-131.}
\lref\borchiv{R. Borcherds, ``The moduli space
of Enriques surfaces and the fake monster Lie
superalgebra,''  preprint (1994).}
\lref\jorgenson{J. Jorgenson and A. Todorov,
``A conjectured analog of Dedekind's eta function
for K3 surfaces,'' Math. Res. Lett. {\bf 2}(1995) 359;
``Enriques surfaces, Analytic discriminants, and
Borcherd's $\Phi$ function,'' Yale preprint. }
\lref\as{
M.F. Atiyah and I.M. Singer,
``Dirac operators coupled to vector bundles,''
Proc. Natl. Acad. Sci. {\bf 81} (1984) 2597}

\Title{ \vbox{\baselineskip12pt\hbox{hep-th/9606082}
\hbox{PUPT-1627}
\hbox{ITEP-TH-18/96}
\hbox{YCTP-P10-96}}}
{\vbox{
\centerline{Chiral Lagrangians,  Anomalies, Supersymmetry,   }
\centerline{and }
\centerline{ Holomorphy}
}}\footnote{}
{ }
\medskip
\centerline{Andrei Losev $^1$, Gregory Moore $^2$,
Nikita Nekrasov $^3$, and Samson Shatashvili $^{4}$\footnote{*}{On leave of
absence from St. Petersburg Steklov Mathematical Institute, St.
Petersburg,
Russia.}}

\vskip 0.3cm
\centerline{$^{1,3}$ Institute of Theoretical and Experimental
Physics,
117259, Moscow, Russia}
\centerline{$^3$ Department of Physics,
Princeton University, Princeton NJ 08544}
\centerline{$^{1,2,4}$ Dept.\ of Physics, Yale University,
New Haven, CT  06520, Box 208120}
\vskip 0.1cm
\centerline{losev@waldzell.physics.yale.edu}
\centerline{moore@castalia.physics.yale.edu}
\centerline{nikita@puhep1.princeton.edu}
\centerline{samson@euler.physics.yale.edu}

\medskip
\noindent
We investigate higher-dimensional
analogues of the $bc$ systems of 2D
RCFT. When
coupled to gauge fields and Beltrami differentials
defining integrable holomorphic structures the
$bc$ partition functions can be explicitly
evaluated using anomalies and holomorphy.
The resulting
induced actions generalize the chiral algebras of
2D RCFT  to
$2n$ dimensions.  Moreover, $bc$ systems
 in four and six dimensions are closely
related to supersymmetric matter. In particular,
we show that $d=4, \CN=2$ hypermultiplets
induce a theory of self-dual Yang-Mills fields
coupled to self-dual gravity. In this way the
$bc$ systems fermionize both the algebraic
sector of the \wzwf\ theory, as defined  by Losev et. al.,
 and the classical open  $\CN_{ws}=2$
string.

\Date{June 11, 1996}

\newsec{Introduction}

This paper continues an investigation into the
application and generalization of two-dimensional
conformal field theory to analogous
higher dimensional theories begun in  \avatar\cocycle
\ref\nthesis{Further results appear in
N. Nekrasov, PhD Thesis}. We will show that
the   theories studied in \avatar\cocycle\nthesis\
are closely related to supersymmetric models.
In particular, we consider actions for gauge
and gravitational fields induced by integrating
over chiral multiplets and hypermultiplets in
four and six dimensions. For   certain
background fields the induced action can be
written explicitly as a ``sigma model
Lagrangian'' in a class of Lagrangians we
name chiral cocycle Lagrangians. The
chiral cocycle
Lagrangians are naturally defined for
all $2n$-dimensional  spacetimes with
a complex structure. Moreover they
are   derivable from characteristic classes.
The chiral cocycle theories  generalize
both the \wzwt\ theory of two dimensions
\wittenwzw\
 (including its gravitational generalization
\ref\polyakov{A.~ M.~ Polyakov,
``Gauge transformations and diffeomorphisms,''
Int. J. Mod. Phys.~ A5: 833, 1990\semi ``Quantum gravity in
two dimensions,''
Mod.~ Phys.~ Lett.~ A2:~ 893,1987;
``Quantum geometry of bosonic strings,''
Phys. Lett. 103B (1981) 207-210} )
and its four dimensional  counterpart
investigated in \ns\avatar. These results are
in  close analogy to the results of Polyakov and Wiegmann and Polyakov
\pwf\plh\polyakov\  in two-dimensions.

To motivate the theories let us begin by recalling
the important notion of two-dimensional
  $bc$ systems  \FMS\knizhnik.
  These are defined by
choosing a complex structure on a   surface
$X_2$ and introducing
anticommuting  $(1,0)$-form and
$(0,0)$-form fields
$b_{m, z}(z,\bar z)$, $c^m(z,\bar z)$,
where $m=1, \dots, r$ is a flavor index.
These fields may be coupled to the
$(0,1)$ part of a gauge field, $\bar a$,
and a Beltrami differential
$\bar \mu_{\zb}^{~~z}(z,\bar z)$, through the
action:
\eqn\bcact{
\int b (\pb + \bar \mu\cdot \p + \bar a) c
}

One can develop all of (rational) conformal
field theory from  the study of the partition function
\eqn\motivate{
\CZ_{ch}(\bar a; \bar \mu) = \int [db dc]
e^{\int b (\pb + \bar \mu\cdot \p + \bar a) c}
=
\bigl\langle e^{\int \bar \mu T + \bar a J} \bigr\rangle
}
The second equation emphasizes the interpretation of
$\CZ_{ch}$ as the generating function of
current correlators. On the other hand,
in two dimensions,  $\bar a$ is (locally)
holomorphically pure gauge, e.g.,
$\bar a = g^{-1} \pb g$ for a $GL(r,\IC)$
gauge transformation $g$. Using the
analytically continued formula for the
anomaly one easily evaluates \motivate,
discovering current algebra from the Lie
algebra cocycle and the \wzwt\ action
from the group cocycle.
 A similar procedure including $\bar \mu$
and the metric
leads to the gravitational $WZW$ theory,
the diffeomorphism anomaly, the
Virasoro algebra, and Liouville theory.

In this paper we indicate how the above
argument generalizes to higher-dimensional
theories. We will define a theory analogous to
\motivate\ in $2n$ dimensions.
For holomorphically trivial fields $\bar a, \bar \mu$
we will use the anomaly to evaluate
$\CZ_{ch}(\bar a, \bar \mu)$ exactly in terms of
a group cocycle.

We generalize \bcact\  to higher dimensions
simply by writing \bcact\ where
$c$ is now a $(0,j)$ form on an $n$-complex dimensional
manifold $X_{2n}$ with values in a vector bundle
$E \rightarrow X_{2n}$,   $b$ is an
$(n,n-1-j)$-form valued in the dual bundle
$E^*$, $\bar a$ is the $(0,1)$ part of a
connection on  $E$ and $\bar \mu$ is a
Beltrami differential.
When  $n$ is larger than one
 an essential new element
emerges: the number of degrees of freedom
in $b,c$ do not balance, and one must introduce
further fields to make sense of \motivate.
If the background fields satisfy:
\eqn\sqrzer{
(\pb + \bar \mu\cdot \p + \bar a)^2
= (\pb_{\bar \mu, \bar a})^2= 0
}
then there is a very natural way to introduce
these extra fields: \sqrzer\ implies that
\bcact\ has a gauge invariance
$ Y: c \rightarrow c + \pb_{\bar\mu,\bar a} \epsilon$.
Gauge-fixing the $Y$-symmetry then  leads
unambiguously to the higher-dimensional
$bc$ theories. We describe this in detail
in section two. The gauge-fixed Lagrangian
has a BRST symmetry $Q$.
It turns out that, on a \k\  manifold
 the $bc$ system is nothing but a Weyl
fermion   (together with a
collection of bosons) coupled to a
gauge field $A= (a,\bar a)$ with both
$(1,0)$ and $(0,1)$ pieces. The
$(1,0)$ piece, $a$, enters through
gauge fixing. In fact, as
we show in section three,  in four and
six dimensions this theory of
bosons and fermions is simply supersymmetric
matter, twisted using a Kahler structure as
in \johansen\park\Witfeb.

The key equation \sqrzer\   has a simple
geometric meaning. The equation for
$\bar \mu$ is the Kodaira-Spencer equation,
so the Beltrami differential
$\bar \mu$ defines an integrable deformation
of complex structure, while  $\bar a$ defines
a holomorphic structure on the vector bundle
$E$ in this complex structure.
The equation \sqrzer\ is solved - locally -
by writing the backgrounds as
``holomorphically pure gauge''
\eqn\puregauge{
\eqalign{
\bar a & = g^{-1} (\pb+ \bar \mu\cdot \p) g\cr
\bar \mu & = \bar \mu[f] = (\pb f)(\p f)^{-1}\cr}
}
where $g(z,\bar z)$ is a $GL(r,\IC)$
gauge transformation, $(z,\bar z)
\rightarrow (f(z, \bar z), \bar z)$ is
a chiral diffeomorphism, and
$(\p f)$ is a matrix of derivatives.
Of course, in general there are
obstructions to writing \puregauge\
globally because
of the existence of moduli.
However,
the dependence on the underlying groups is
of primary importance and is hence the focus of
this paper. Thus, in this paper we ignore
moduli and zeromodes and focus on the $g$ and
 $f$ dependence.
The very interesting and important
problem of dependence on moduli is a
subject to which we hope to return.

In section four we address the definition of the
chiral partition function \motivate, and in
section five we evaluate it.
In direct analogy to the two-dimensional
case the partition function \motivate\
can be evaluated exactly for the class
of backgrounds \sqrzer. The essential
argument for this is extremely simple and
proceeds as follows. Since we have a Weyl
fermion the chiral partition function must
have an anomalous variation under
unitary gauge transformations of $A$.
In an appropriate regularization
(which we will refer to as the
``standard regularization'')
\eqn\anom{
Z_{ch,s}(A^u) = e^{i \alpha_s(A,u)} Z_{ch,s}(A)
}
where $ \alpha_s(A,u)$ is the standard
anomaly functional
derived from descent \fs\zumino\anomrev.
\foot{In more algebraic language, $\alpha_s(A,u)$ is
a ``1-cocycle taking values in functions of
connections.'' It satisfies $\delta \alpha_s =0$
where $(\delta f)(A,g) = f(A^g )-f(A) $ for
0-cochains, $(\delta f)(A,g_1, g_2) = f(A^{g_1}, g_2) -
f(A, g_1 g_2) + f(A ,g_1) $ for 1-cochains,
etc. }
On the other hand, if $\CZ_{ch}$ can
be defined maintaining $Q$-invariance
then it must be a   function
of $\bar a$ only, that is, it must be $a$-independent.
(More generally it must be   independent
of gauge fixing):
\eqn\holo{
\CZ_{ch}(A) = \CZ_{ch}(\bar a)\quad .
}
A glance at the standard formula for
$\alpha_s$ (see, e.g., eq. $(5.2)$ below)
shows that $Z_{ch,s}$ cannot be holomorphic,
hence the standard regularization does not
preserve $Q$-symmetry.
The situation is redeemed by the addition of
a local counterterm, or, what is the same, by
a change of regularization.
 In more algebraic langauge, a change of
regularization corresponds to a change of the
anomalous
group 1-cocycle $\alpha_s$ by a coboundary
$\alpha_s \rightarrow \alpha= \alpha_s + \dge \gamma$.
The absence of a $Y$-anomaly implies the
preservation of $Q$-invariance.  A necessary
condition for $Q$ invariance is
the existence of a local functional
$\gamma(a,\bar a)$ such that the group cocycle
\eqn\triv{
\alpha_h(\bar a, u)= \alpha_s(A,u) + \dge \gamma(a,\bar a,u)
}
is $a$-independent. In section 5.1 below we show that
such a $\gamma$ exists. After analytic
continuation of \anom\
in $u$, we are in a position
to gauge away $\bar a= g^{-1} \pb g$
from $\CZ_{ch}$. The result is
the  chiral cocycle Lagrangian:
$\Gamma[g]= {i\alpha_{h}(\bar a=0 ,u=g)}$
where $g$ is in the complexified gauge
group. It is possible to write explicit
formulae for $\alpha_h$. The Lie algebraic
cocycle corresponding to \triv\ is very simply written:
\eqn\laco{
C(\epsilon, \bar a) =
\int_{X_{2n}} \Tr \epsilon(\p \bar a)^n
}
and the corresponding chiral cocycle Lagrangian
(in four dimensions) is:
\eqn\rsngr{
\log \CZ_{ch}[\bar a] =
\Gamma[g]  = \int_{X_4} \Tr\Biggl[
2(\ell \bar \ell - \bar \ell   \ell ) \pb \ell + (\bar \ell   \ell)^2 \Biggr]
+ {2 \over  5} \int_{X_4\times I} \Tr (g^{-1} dg)^5
}
where $\ell=g^{-1}\p g, \bar \ell = g^{-1}\pb g$.
 As we show in section
seven,  $\Gamma[g]$
can be obtained directly from
a holomorphic
version of the descent procedure \fs\zumino\anomrev
and has the form
\eqn\gencocy{
\Gamma[g] =
\int_{X_{2n} \times D}
 {dw \over  w} \wedge \Tr \bar \ell (\p \bar \ell)^n
}
where $D$ is a disk and
we extend $g$  so that $g=1$ on the boundary
of $g$.

Studying the  $\bar \mu$-dependence
of \motivate\ in a similar spirit leads to
analogous formulae for
group cocycles for the group of
chiral diffeomorphisms.  Parametrizing $\bar \mu$
as in  \puregauge\ we find in section 8  that
$\log \CZ_{ch}[\bar \mu] = \Gamma[J(F)^{-1}]$ where
$F$ is an inverse to $f$ and $J(F)$ is the
$GL(n,\IC)$ matrix $\p_i F^j$, i.e. the
chiral Jacobian. The Lie algebra cocycle
analogous to \laco\ is simply:
\eqn\diffcob{
 C(v,\bar \mu)  =  \int_{X_{2n}} \Tr\biggl[
(\p v) (\CR_{\bar \mu})^{n} \biggr]
}
where  we define
a two-form $\CR_{\bar \mu}$ with values in
$gl(n,\IC)$ by
\eqn\mutoo{
\eqalign{
(\CR_{\bar \mu})_m^{~~p} & \equiv dz^j d\zb^\kb
\bigl( \p_j \p_m \bar \mu_{\kb}^{~p} \bigr)\cr}
}
and the trace is in any representation.
Given \motivate, \diffcob\
generalizes the Virasoro anomaly to higher
dimensions.

There are some noteworthy implications
of the above
simple argument. First, the principle of holomorphy
determines the kinetic term in the induced
``pion Lagrangian.'' Usually such kinetic terms
(and their higher-derivative corrections) are
regarded as   coboundaries which
definitely affect
physics but are beyond the reach of the method
of descent. Evidently,
a principle analogous to our holomorphy
principle in the context of  QCD would have
profound consequences
\foot{There are other possible applications to
QCD. See, for example,    \bardeen\cangemi. }
Second,  quantization of free field theory
is in principle straightforward, so
we have provided a construction of
higher dimensional counterparts of chiral
algebras (in the sense of CFT), together with
a realization  of these algebras generalizing
two-dimensional fermionization.
\foot{In the special case that
$X_4$ is a product of Riemann surfaces
on can define ``projected algebras'' which
are exactly the chiral algebras of 2d CFT.
These were first investigated in
\johansen\ and further studied in
\avatar. Related results on higher-dimensional
fermionization are described in
\banks. }

In section nine we  consider the induced action for a four-dimensional
$bc$ system in a special representation
defined by  an $\CN=2$ hypermultiplet. In this
representation the action \rsngr\ includes the
\wzwf\ Lagrangian \avatar. (Indeed this was
the starting point for the present investigation.)
The induced theory is also closely related to the
classical target space theory of an open
string theory with gauged $\CN_{ws}=2$ supersymmetry.
 In particular the Plebanski equation and its modification
in the presence of gauge fields naturally emerges
from the chiral cocycle Lagrangian.
Our results suggest many new questions
and interesting avenues of research. Some of these
are indicated in the conclusions.

\newsec{Lagrangians for
$bc$ systems in $2n$ dimensions}

In this section we define more precisely
the Lagrangians of
the higher-dimensional analogs of
the 2d $bc$ systems of \FMS\knizhnik.

\subsec{$bc$ systems in $4$ dimensions}

Let $E\to X$ be a complex rank $r$ hermitian
vector bundle over a complex surface $X$.
We assume that $E$ is equipped with
a $GL(r,\IC)$ connection  $(a,\bar a)$  so that
the $(0,1)$ piece $\pab$ defines a holomorphic
structure on $E$: $\pab^2=0$. Note that we
do {\it not} assume $F^{2,0}=0$.

The most obvious generalization
of bc systems is defined by introducing the
fermionic fields
\eqn\bci{
\eqalign{
b & \in \Omega^{2,0} (X, E^*) \cr
c & \in \Omega^{0,1} (X, E ) \cr}
}
We try the action:
\foot{ The symbol
$\langle\cdot, \cdot \rangle$ denotes a bilinear,
dual pairing, while $(\cdot, \cdot )$ will denote
a Hilbert space inner product, antilinear in the
first slot. }
\eqn\bcii{
S {\buildrel ? \over =}
 \int_X \langle b, \pab c \rangle
}
This  theory does not makes sense:
the operator   $\pab$ is not nearly
Fredholm. Indeed, there is a large
 gauge symmetry:
\eqn\bciii{
c \rightarrow c + \pab \epsilon
}
which we refer to as $Y$-symmetry.
We now fix it using the Faddeev-Popov procedure.
Introduce a ghost $\phi$ and an
antighost multiplet $(e,\bar\phi)$:
\eqn\bcvi{
\eqalign{
\phi & \in \Omega^{0,0} (E) \cr
e,\bar\phi & \in \Omega^{0,0}  (E^* ) \cr}
}
(fields with Greek letters commute, Latin letters
anticommute.) We have  a BRST $Q$-symmetry:
\eqn\bcvii{
\eqalign{
Q: c  \rightarrow  c + \pab \phi\quad & \quad \phi \rightarrow 0 \cr
Q:  \bar\phi  \rightarrow e \quad & \quad   e  \rightarrow 0\cr}
}

We introduce the
  gauge-fixing term via the gauge fermion:
\eqn\bcxxi{
\Psi  = -i \int_X \omega
\langle \p_a \bar\phi ,  c \rangle
}
where $\omega$ is a positive $(1,1)$
form. For example we can take it to be
the imaginary part of an hermitian metric on
$X$.

The action of the gauge-fixed chiral bc system is
\eqn\bcviii{
\eqalign{
S_{bc} & =  \int_X \langle b, \pab c \rangle + \{Q, \Psi\}\cr
& = \int_X \langle b, \pab c \rangle+
\omega \langle \p_a  e,   c \rangle +
\omega \langle  \p_a \bar\phi,   \pab \phi \rangle\cr}
}
The equations of motion are:
\eqn\bcxxx{
\eqalign{
\pab c= \p_a(\omega c) & = 0 \cr
\omega \p_a e - \pab b & =0 \cr
 \p_a \omega \pab \phi& = 0 \cr
 \pab \omega \p_a \bar\phi& = 0 \cr}
}
The explicit coupling to the
gauge fields is:
\eqn\bcx{
- \int \Tr\biggl( \bar a J + a\bar J
\biggr)
}
where
\eqn\bcxi{
\eqalign{
J & =    c \otimes b - \omega  \phi \otimes \p_a \bar\phi \in
\Omega^{2,1}(End E)\cr
\bar J & = \omega  c \otimes e
+ \omega \pb
 \phi \otimes \bar \phi \in \Omega^{1,2}(End E)\cr}
}
In particular using the equations of motion we have:
\eqn\bcxii{
\{ Q, J\} = - \pab (b \phi)
}
and
\eqn\bcxii{
\pab J =
\{ Q, \omega \p_a \bar \phi c\}
}

Of crucial importance in this paper is the
fact, following immediately from \bcviii, that
 the coupling to $a$ is $Q$-exact:
\eqn\bcxxii{
{\delta \over  \delta a} S =
\bar J  + \omega \bar a \phi \otimes \bar \phi =
\{ Q, \omega  c \bar\phi \}
}
and hence the path integral must
be  $a$-independent. This   is
simply an example of the independence
of gauge slice in Faddeev-Popov gauge
fixing. More generally, one should be able
to obtain equivalent results using different
gauge fermions in \bcviii.
The same arguments suggest
that that the dependence on the
form $\omega$
is likewise $Q$-exact.
This is more
subtle, since $\omega$ can appear
in the measure. We will return to this issue in
section 7.2 below.

It is straightforward to
introduce a coupling to a Beltrami differential
coupling to $\int \bar \mu T$ with
\eqn\energmom{
\eqalign{
 T & =   \langle b {\buildrel \otimes \over  ,}  \p c
\rangle + \langle
 \p(\omega \bar \phi) {\buildrel \otimes \over  ,} \p \phi \rangle
\in \Omega^{2,1}(T^{1,0*} X) \cr}
}
$T$ is essentially the stress-energy tensor with
the volume form included. Since we have
included the volume form we have index structure:
\eqn\emtensi{
T_{ij\bar k \ell} = b_{ij} \p_\ell c_{\bar k}
+ \p_{[i}(\omega_{j ]\bar k} \bar \phi) \p_\ell \phi
}

Remarks:

\item{1.} In an entirely similar way, if
$F^{2,0}(a)=0$, but $\bar a$ is arbitrary we can
gauge fix the action
\eqn\otheract{
\int
\omega \langle   \p_a e, c \rangle
}
The scalars become $(2,0),(0,2)$-forms and the
current coupling to $\bar a$ is $Q$-exact.

\item{2.} Henceforth in this paper for simplicity
we assume that $X$ is a K\"ahler manifold. This
allows us to use the K\"ahler identities
$\pb^\dagger = [\Lambda, \p]$ where $\Lambda$ is
the contraction with the K\"ahler form. In section four
  we will integrate $\p_a$ by parts and write
formulae in terms of $(\pb)^\dagger_a$.

\subsec{Generalization to $2n$ dimensions}

The generalization to $2n$ dimensions is
straightforward. It is easiest to work
on a K\"ahler manifold.
The gauge fixing procedure
now produces a sequence of ghosts for
ghosts. Schematically the Lagrangian has the
form
$$
\int b (\pb + \pb^\dagger ) c
+ \tilde \phi \pb\pb^\dagger \phi+ ...
$$
Here $b\in \oplus \Omega^{2p,0}(E^*)$
while $c\in \Omega^{0,2p-1}(E)$ (and
for simplicity we have
omitted powers of the Kahler form $\omega$).
As we recall below the first term is simply the
action for a Weyl fermion in a particular background
gauge field.

\newsec{$bc$ systems as supersymmetric field theories}

\subsec{Chiralmultiplets and vectormultiplets in $d=4 $}

On a
K\"ahler surface $X$  we can
consider a chiral superfield $\Phi$ and twist
it by a half-integral twist with the $R$-current
 \johansen\park\Witfeb. \foot{We work in Euclidean space so fermions are
Weyl fermions.}
The result is that
 we can
turn the Weyl fermion into anticommuting
form fields $\psi_\alpha \rightarrow c$,
$\bar \psi^{\dot \alpha} \rightarrow b,e$.
Put more mathematically,
we consider a chiral superfield $\Phi$ in the
bundle $E\otimes K^{-1/2} $. The conjugate
superfield $\bar \Phi$ is in $E^*\otimes K^{+1/2} $,
so, by a standard isomorphism,
the fermions in the theory are sections:
\eqn\twi{
\eqalign{
 \psi_{\alpha} & \in \Gamma[S^- \otimes K^{-1/2} \otimes E]
\cong \Omega^{0,1}(E) \cr
\bar \psi^{\dot \alpha}
& \in \Gamma[S^+ \otimes K^{1/2} \otimes E^*]
\cong \Omega^{0,0}(E^*) \oplus \Omega^{2,0}(E^*) \cr}
}
Thus from a free superfield
we get exactly the field content of the
$bc$ system. Moreover, under the
isomorphism \twi\
the Dirac operator with spinor covariant
derivative coupled to $K_X^{\pm 1/2}$:
\eqn\drccov{
\eqalign{
\nabla_\mu^- & = \p_\mu - {1 \over  2}\omega_\mu^{ab}
(   \sigma^{ab})
\pm  V_\mu+ A_\mu \cr
\nabla_\mu^+ &  = \p_\mu - {1 \over  2}\omega_\mu^{ab}
( \bar  \sigma^{ab})
\pm V_\mu + A_\mu \cr}
}
where $V_\mu=  \p_\mu\log \sqrt{g}$
is mapped to the Dolbeault operator according
to
$\bar \sigma\cdot \nabla^-
 \rightarrow \pab \oplus  \pab^\dagger$
for $K^{-1/2}_X$, etc.
Thus coupling to  $K^{-1/2}$ is equivalent to
topological twisting by the $R$-current.
The action:
\eqn\freesus{
\int \sqrt{g} \biggl[ g^{\mu\nu} \langle \nabla_\mu \bar \phi,
\nabla_\nu \phi \rangle + i \langle \bar \psi_{\dot \alpha} ,
\bar \sigma^{\mu \dot \alpha \alpha} \nabla_\mu^- \psi_\alpha \rangle
  \biggr]
}
is equivalent to \bcviii\ if
$f\equiv g^{i \bar j} F_{i \bar j} =0$. More generally, if we
consider the coupled vectormultiplet-chiralmultiplet
system in a vectormultiplet
background with $\lambda=F^{0,2}=0$ (where
$\lambda$ is the gluino)
the standard Lagrangian and the $bc$ Lagrangian
differ by $Q$-exact terms.
Finally the supercurrent
$\bar q^{\dot \alpha,\mu} = \langle
\bar \sigma^\nu \sigma^\mu \bar \psi, \nabla_\nu \phi\rangle$
is conserved when the
background SYM fields are antiself-dual.
Using the projectively covariantly
constant spinor on a Kahler manifold
 we can produce the operator $Q$ from $\bqa$:
\eqn\sprchr{
Q =  \int_{X_3}  \bar s_{\dot \alpha}
\bar q^{\dot \alpha \mu} n_\mu
}
Even if $f\not=0$ $Q$ is still a conserved
charge as long as $F^{0,2}=0$.
We now recognize the
$bc$ system as a twisted Euclidean
chiral supermultiplet
in a supersymmetry-preserving background.

One may also consider
the  $bc$ system corresponding to
\otheract.
Its partition function turns out to
be related to that of a
vectormultiplet, expanded in background field
gauge.

\subsec{Hypermultiplets and vectormultiplets in $d=6$}

The Lagrangians for hypermultiplets
and vectormultiplets on K\"ahler manifolds
 in six dimensions  can be
related to $bc$ systems along the above lines.

In $d=6, \CN=(1,0)$  supersymmetry a
hypermultiplet consists of Bose/Fermi
fields $\Phi $ and $ \psi$
transforming in in the $SU(4) \times USp(2)_{\CR}$
Lorentz $\times$ R-symmetry group as
  $(1;2)$   and $  (4; 1) $ respectively. On a Kahler manifold
the holonomy group is $U(3)$. Twisting by
a linear combination of the $U(1)$ in $U(3)$ and
the internal symmetries we produce the Lagrangian
\eqn\hyper{
\eqalign{
I  =
\int_{X_6} \omega b^{2,0}\pb c^{0,1} +
&
\omega \langle b^{2,0} , \pb^\dagger c^{0,3} \rangle
+ \omega^3 \langle b^{0,0},
\pb^\dagger c^{0,1}\rangle
\cr
+  \int_{X_6} \omega^3 \langle \tilde \phi^{0,0},
\pb^\dagger \pb \phi^{0,0} \rangle
&
+ \langle \p \p^\dagger \bar \phi^{3,0}, \phi^{0,3}\rangle
\cr}
}
where superscripts indicate the form type.

By choosing an appropriate twisting we have
a scalar $Q$-symmetry:
\eqn\scalsus{
\eqalign{
Q:  c^{0,1}  \rightarrow   \pb \phi^{0,0} &  \qquad
\phi^{0,0} \rightarrow 0 \cr
 \bar \phi^{0,0} \rightarrow - b^{0,0}
&\qquad  b^{0,0} \rightarrow 0 \cr
 b^{2,0} \rightarrow   \p^\dagger \bar \phi^{3,0}
& \qquad \bar \phi^{3,0} \rightarrow 0 \cr
 \phi^{0,3} \rightarrow c^{0,3}
& \qquad c^{0,3} \rightarrow 0 \cr}
}
such that the action is:
\eqn\qexact{
I = \int \omega b^{2,0} \pb c^{0,1} +
Q\biggl( \int \omega^2 \p \bar \phi^{0,0} c^{0,1} +
\p b^{2,0} \phi^{0,3} \biggr)  }

In much the same way, by
gauge fixing
$$
\int b^{3,0}\pb c^{0,2}
$$
we produce the Lagrangian used for evaluating
a $d=6, \CN=(1,0)$
vectormultiplet in background field gauge.

\subsec{d=2n}

On a Kahler manifold there is a standard relation
between the Dirac operator coupled to a vector bundle
$E\otimes K_X^{\pm1/2}$ and the Dolbeault operator
coupled to $E$ \BGV.
Given an
Hermitian vector bundle on an arbitrary
Kahler manifold:
\eqn\genkhlr{
\eqalign{
\Gamma[S^+ \otimes K^{1/2}\otimes E] & \cong \oplus \Omega^{2p,0}(E)\cr
\Gamma[S^- \otimes K^{1/2}\otimes E] & \cong \oplus \Omega^{2p-1,0}(E)\cr
\Gamma[S^+ \otimes K^{-1/2}\otimes E] & \cong \oplus \Omega^{0,2p}(E)\cr
\Gamma[S^- \otimes K^{-1/2}\otimes E] & \cong \oplus \Omega^{0,2p-1}(E)\cr
}
}

Under this isomorphism the Weyl operator becomes the
Dolbeault operator according to
 \eqn\dirdoli{
\matrix{
\Gamma[S^+ \otimes K^{-1/2}\otimes E] &
{\buildrel D^+ \over  \rightarrow} &
\Gamma[S^- \otimes K^{-1/2}\otimes E]\cr
\downarrow &  & \downarrow\cr
\oplus \Omega^{0,2p}(E)
&
{\buildrel \pab \oplus  \pab^\dagger
 \over  \rightarrow} & \oplus \Omega^{0,2p-1,0}(E)
\cr}
}

Accordingly, the physical content of all
higher-dimensional $bc$ systems consists of
a twisted Weyl fermion together with a set of
  $(0,p)$-form bosons.
When coupled to a nontrivial vector bundle
these theories are
(noninteracting!) nonabelian
$p$-form theories.

\newsec{Defining the chiral partition function}

\subsec{ Perturbation theory point of view }

The most straightforward definition of the path integral
of the $bc$ system
is as the generating functional for current
correlation functions:
\eqn\currcorr{
Z_{ch,s}[a, \bar a, \mu, \bar \mu] =
\biggl\langle \exp \int \Tr(\bar a J) + \bar \mu T \biggr\rangle
}
defined by perturbation theory in $\bar a, \bar \mu$.
Since the theory is simply a theory of a
Weyl fermion together with a collection of
bosons we can use the standard perturbation
series of quantum field theory.
The propagators can be defined, for example,  by introducing
a Dirac fermion with left and right currents coupled
to different gauge fields.  In the present case we
  expand in $\bar a, \bar \mu$ and
use the free propagators:
\eqn\freeprops{
\eqalign{
\langle \phi^m(x) \omega(y)\p_{ y} \bar \phi_n(y) \rangle
& = {i \delta^m_{~~n}\over  4 \pi^2}
  { (\bar x-\bar y)^\ib \epsilon_{\ib\jb} d \bar y^\jb \over  (x-y)^4}dy^1 dy^2
\cr
\langle c^m(x) b_n(y) \rangle
= {\delta^m_{~~n} \over    \pi^2}  &
{ (\bar x-\bar y)^\ib \epsilon_{\ib\jb} d \bar x^\jb \over (x-y)^4} dy^1 dy^2
\cr}
}
This procedure plainly produces a holomorphic
function of $\bar a, \bar \mu$.
Of course, this procedure also produces infinities
  from divergences at coincident points.
Thus the chiral perturbation series in fact depends
on extra data, as is standard in quantum field theory.
For example, defining the short-distance expansion
of a Green's function requires comparing two fibers
of a vector bundle at nearby, but different points.
It is thus natural to introduce a complete connection
$A= (a, \bar a)$ when subtracting infinities.

Because of supersymmetric cancellations the leading
term in the expansion is cubic in $A$. The resulting
functional is of course anomalous under gauge transformations
because of the Weyl fermion. The bosons only
contribute to the anomaly through a coboundary.
In what follows we assume that the theory
can be quantized   with
the standard formula for the gauge anomaly of
Weyl fermions \fs\zumino\anomrev.

\subsec{Factoring Determinants point of view}

The partition function \currcorr\ is formally evaluated
by writing out the free field modes.:
 \eqn\chrlpf{
\eqalign{
\CZ_{ch}
 ~ {\buildrel ? \over =} ~
 \int [d e d b d c] [d\phi d \tilde \phi]
\exp \biggl\{ - S_{bc}  \biggr\}  = &
{ \Det' [( \pab \oplus(  \pb  )^\dagger_a)]_{(0,1)}
\over
\Det' [(  \pb  )^\dagger_a \pab  ]_{0,0} } \cr}
}
in four dimensions, with a simple generalization to
$2n$-dimensions. In this paper we ignore
moduli
so zeromode insertions are not needed.
As we have emphasized throughout the paper,
if there is no $Y$-anomaly then in fact
$\CZ_{ch}$ is only a function of $\bar a$.

The expression \chrlpf\ is formal.
Any definition using determinants must face up to the
fact that $\pb + \pb^\dagger$ is an
operator between different spaces.
The time-honored approach to this
difficulty is the introduction of
a system of the opposite
chirality: $\bar b\in \Omega^{0,2}(E),
\bar c \in \Omega^{1,0}(E^*)$ etc. In terms of
the Weyl fermions we are introducing the
complex conjugate representation. We can
define a measure on the resulting theory.
\foot{
For a real representation on
a  hyperkahler manifold   we could
define a mass term using the nowhere-vanishing
$(2,0)$ form. }
Coupling the opposite
chirality to the gauge fields $(b, \bar b)$
we obtain  the general nonchiral system:
\eqn\gennonch{
\IZ[a, \bar a; b, \bar b] \equiv
{\Det [(\pb_{\bar b} + (\pb^\dagger)_b) (\pb_{\bar a} + (\pb^\dagger)_a)
]_{0,1}
\over
\Det [(\pb^\dagger)_{a} \pb_{\bar a} ]_{0,0}
\Det [(\pb^\dagger)_{b} \pb_{\bar b} ]_{0,0} }
}

{}From the Q-symmetry of the classical action for
nonchiral system
(provided that $F^{(0,2)}(\bar a)=F^{(2,0)}(b)=0$)
we conclude that
\eqn\logzee{
\log \IZ[a, \bar a; b, \bar b]
= \log\CZ_{\rm nonchiral} [\bar a, b]+ \CC_{\rm local}
[a, \bar a; b, \bar b]
}
where $\CC_{\rm local}[a, \bar a; b, \bar b]$  is a local
counterterm.
Note that we do not impose any conditions on
gauge fixing fields $ a , \bar b $.

In order to compute
nonchiral function $\CZ_{\rm nonchiral} [\bar a, b]$ we will take
$a=b$, and $\bar b= \bar a$, so that we get, up to
local counterterms:

\eqn\nonchirsub{
\CZ_{\rm nonchiral}[\bar a, b] =
{\Det [ \{ (\pb^\dagger)_{b}, \pb_{ \bar a} \} ]_{0,1}
\over
\bigl(\Det [ (\pb^\dagger)_{b} \pb_{\bar a} ]_{0,0}\bigr)^2 }
}

 The right hand side of this equation is nothing but
Ray-Singer (RS)-torsion, and the problem of computation of
chiral part of the partition functon is therefore
just the problem of
factorization of RS-torsion into a product of
terms depending only on $\bar a$ and $b$ up to local counterterm:
\eqn\nonchirsub{
\CZ_{\rm nonchiral}[\bar a, b]  = \CZ_{ch}(\bar a)
\bar \CZ_{ch}(b) \exp \CC_{\rm local}[\bar a, b]
}

We will study the generalized RS-torsions
(containing $\mu$) their factorizations and
corresponding gauge and
diffeomorphism anomalies of chiral systems
 with the help of supersymmetric
quantum mechnanics in the next section.

\subsubsec{Determinant lines and Quillen norms}

There is a well-developed mathematical theory
of the objects \chrlpf: this is the theory of
determinant lines and Quillen metrics.
$\CZ_{ch}$ should be regarded as a
section of  a determinant line bundle $\lambda$ over a
parameter space $\CS$. In this case $\lambda$ is dual to the
line bundle
\eqn\dtln{
\Lambda^{mx} H^{0,0}_{\pb}(X;E) \otimes
(\Lambda^{mx} H^{0,1}_{\pb}(X;E))^{-1} \otimes
\Lambda^{mx} H^{0,2}_{\pb}(X;E)
}
where $\Lambda^{mx}$ is the maximum
exterior power.
 While $\lambda$ is canonically trivial
(absence of zeromodes) it carries a highly
nontrivial
Quillen metric:
\eqn\qllnrmi{
\parallel \CZ_{ch} \parallel^2\equiv
 {\Det' [(\pab  )^\dagger \pab  ]_{0,1}
\over
\Det' [(\pab  )^\dagger\pab]_{0,0}  }
}
which is again RS-torsion \BGS.
 The zeromodes are easily
reinstated in the above formulae.
The above remarks generalize straightforwardly
to the $2n$-dimensional case.

It turns out, as we will demonstrate in the
next section that the nonchiral
partition function can be split holomorphically:
\eqn\holosplitt{
\CZ_{\rm nonchiral}[\bar a, a] =
\parallel \CZ_{ch} \parallel^2 = e^{\Gamma[a,\mu] + K(a,\bar a; \mu, \bar \mu)
+
\Gamma(\bar a, \bar \mu) }
}
where $\Gamma$ is a nonlocal, but holomorphic functional,
while $K$ is a local, but nonholomorphic functional.
$\exp\Gamma$ defines the chiral partition function
$\CZ_{ch}$.

\newsec{Three derivations of
the chiral partition function}

In this section we describe three
methods to evaluate the
partition function for $\bar a, \bar \mu$
satisfying \sqrzer.

\subsec{Anomaly plus holomorphy}

For simplicity we work in four dimensions,
  put $\bar \mu = 0 $,
and consider the definition of the chiral
partition function via the perturbation expansion.
The process of gauge fixing introduces a
$(1,0)$ connection $a$. Let us assume that the
perturbation expansion \currcorr\
can be regularized so that
under unitary gauge transformations we have the
standard anomalous variation:
\eqn\ccc{
Z_{ch,s}(a^g, \bar a^g)
= e^{i \alpha_s(a,\bar a, g)} Z_{ch,s}(a, \bar a)
}
for a cocycle $\alpha_s$ derived from
the descent procedure. Explicitly, in
4-dimensions we have the standard formula
\eqn\can{
\eqalign{
\alpha_s(\bar a, a, g) &={1\over  240 \pi^2}
\int_{X_4\times I}
\Tr(g^{-1}dg)^5 \cr
-{1 \over  48 \pi^2}
&
{\int_{X_4} }\Tr[(AdA + dA A +A^3)dgg^{-1}
- (Adgg^{-1})^2 - A(dgg^{-1})^3] \quad .\cr}
}
where $A=a+\bar a$ and
with the corresponding Lie algebra cocycle
( for $g=1+\epsilon+\cdots $) given by:
$\int \Tr \epsilon[2 (d A)^2 + d(A^3) ] $.
Evidently, the standard cocycle depends
on both $a$ and $\bar a$, so the standard
regularization violates BRST decoupling.
Nevertheless, one can introduce the
counterterm:
\eqn\frstct{
\gamma(a,\bar a) =
\int \Tr[(2 \p \bar a + \pb a)(a \bar a - \bar a a)
+ {3 \over  2} (a \bar a)^2 ]
}
and  check, by direct calculation, that
\eqn\holoct{
 \int \Tr \epsilon[2 (d A)^2 + d(A^3) ]
-
\delta_\epsilon\gamma(a,\bar a)
= 6 \int \Tr \epsilon (\p \bar a)^2
}
Therefore the chiral partition function
$\CZ_{ch}(\bar a)
\equiv  e^{i \gamma(a,\bar a)} \CZ_{ch,s}(a,\bar a)$
satisfies:
\eqn\anomm{  \CZ_{ch}(\bar a^g)=e^{i\alpha_h(\bar a, g)}{  \CZ_{ch}(\bar a)}}
with the group cocycle:

\eqn\new{\alpha_h(\bar a,g) = \alpha_{s}(a,\bar a, g)
+ \dge \gamma =
\alpha_{s}(a,\bar a, g) + (\gamma(a^g, {\bar a}^g) - \gamma(a, \bar a))}

Now we can analytically continue formula \anomm\ from unitary to complex
transformations   and  (locally)
gauge away $\bar a$: $\bar a = g^{-1}{\bar \partial}g$. Thus
the $\bar a$-dependence of the chiral
partition function is given exactly by the effective
action:

\eqn\main{\CZ_{ch}(\bar a)
= e^{\Gamma[\bar a]} = e^{i\alpha_{h}( 0 ,
g)}}

This is a central result of this paper.
Later we will give two more derivations
of this formula and will also include gravity
(via the diffeomorphism dependence of the
chiral partition function).

Recall that in two-dimensions the
  trivial cocycle is  also  fixed by holomorphy
and leads to the kinetic term in the \wzwt\ action.
In 4d there is an analogous term but
it is of dimension four. An ambiguous term
$\int \Tr \bar a \alpha $ remains, where
$\alpha$ is a $2,1$ form. (For example,
it could be that $\alpha \sim \omega \p K_0$
for the background Kahler metric.)

Remark: On general grounds one expects that
in $2n$ dimensions there is a local
counterterm $\gamma(a,\bar a)$ satisfying
\new. This counterterm should be closely
related to  the ``Bardeen counterterm''
relating VA and LR forms of the anomaly
\anomrev. In any case, the explicit form
of $\gamma(a,\bar a)$ is not needed for
the construction of the chiral cocycle
Lagrangians below.

\subsec{Family index theorem and the Quillen anomaly}

Let us now recall
the topological interpretation of anomalies
following from the family index theorem
\as\BGV\phys. The (differential form version) of the
family index theorem
 has already been applied in this context in
\doniii\donii\ and, most definitively, by
Bismut, Gill\'e and Soulet \BGS.
The determinant line bundle $\lambda \rightarrow \CS$
over a parameter space $\CS$
has first Chern class:
\eqn\curv{
\p^{(\CS)} \pb^{(\CS)}  \log \parallel \CZ_{ch} \parallel^2
= (2 \pi i )[\int_X \Td(T^{1,0}Z ) \ch(\hat E)]^{(2)}
}
where $Z \rightarrow \CS$ is
fibered by $X$, $T^{1,0}Z$ is the
relative tangent bundle,
 and $\hat E$ is the universal
bundle. In fact, using $\zeta$-function
regularization we may choose the
Bismut-Freed connection on $\lambda$ and
obtain an explicit differential form
representative of the RHS of \curv.
When the parameter space is a quotient
by a gauge group e.g. $\CS= \CA/\CG$ then
the line bundle $\lambda$ descends
from a $\CG$-equivariant line bundle
on $\CA$.
The curvature is directly related to the
equivariance of $Z(a,\bar a)$ under gauge
transformations.

One may explicitly evaluate
the RHS of \curv\ using the techniques
explained in the next section. Putting
$R=0$ for simplicity we find, by direct
computation, that  \curv\ has the form
$\p^{(\CS)} \pb^{(\CS)}  \beta(a, \bar a)$
where
\eqn\frmbet{
\eqalign{
\beta(a, \bar a) & = {1\over  2 \pi}
\int_{X_2} \Tr a \bar a \qquad \qquad n=1 \cr
  & =  {1 \over  8 \pi^2}  \int_{X_4} \Tr\Biggl( \bar a a \pb a - a \bar a \p
\bar a
+ \half (\bar a a)^2  \Biggr)  \qquad n=2 \cr}
}

Thus we have justified \holosplitt\ with
$K(a,\bar a) = \beta(a, \bar a)$.
Of course, the full partition function is
(unitarily) gauge invariant while $\beta$ is not.
Indeed from the the gauge transformations
of $\beta$ one can derive the holomorphic
action $\Gamma[\bar a] $:
\eqn\gaugek{
\dge \beta(a,  \bar a, g) =
-  \delta  \bar \Gamma[a,g]
- \delta   \Gamma[\bar a,g]
}
and hence the derive
chiral partition function. We will describe the
formula in the next section.

\subsec{Derivation from supersymmetric quantum
mechanics}

One rather direct way to prove the index theorems
uses supersymmetric quantum mechanics
\luis\fw. Using the techniques used to prove
the index theorems one can
also study the Lie algebra cocycles. This
method is very simple and, in the case of
diffeomorphism cocycles is the most direct
route to the answer.

As is well-known Ray-Singer torsion can
be defined using a trace in a supersymmetric
quantum mechanics system with bosonic
coordinates $\phi^i, \bar \phi^{\ib}$ and
corresponding fermionic coordinates
$\psi^i, \bar \psi^{\ib}$.
Suppose we have two $\pb$-operators:
$(\pb + \bar \mu_1 \cdot \p + \bar a_1 )^2=0$,
and
$(\pb + \bar \mu_2 \cdot \p + \bar a_2 )^2=0$.
Introducing a  reference
Kahler metric $g^{(0)}$ we can take
$Q_1=\pb + \bar \mu_1 \cdot \p + \bar a_1 $
and $\bar Q_2
= (\pb + \bar \mu_2 \cdot \p + \bar a_2)^\dagger$
as right and left arrows respectively in the
complexes:
\eqn\cplx{
\eqalign{
0 \longrightarrow \Omega^{0,0}
{}~ {\buildrel
Q_1 \over  \longrightarrow} ~
\Omega^{0,1}
&
{}~ {\buildrel
Q_1 \over  \longrightarrow} ~
\Omega^{0,2}
{}~ {\buildrel
Q_1 \over  \longrightarrow} ~
 \cdots \Omega^{0,n} \longrightarrow 0 \cr
0 \longleftarrow \Omega^{0,0}
{}~{\buildrel
\bar Q_2 \over  \longleftarrow} ~
\Omega^{0,1}
&
{}~{\buildrel
\bar Q_2 \over  \longleftarrow} ~
\Omega^{0,2}
{}~{\buildrel
\bar Q_2 \over  \longleftarrow} ~
 \cdots \Omega^{0,n} \longleftarrow 0 \cr}
}

This complex can be realized in the
context of supersymmetric quantum mechanics:
$\oplus \Omega^{0,*}$ is the Hilbert space of
states,
$Q_1, \bar Q_2$ are supersymmetry operators,
there is a fermion number operator $F$ such that
$\{ F, Q_1\} = Q_1, \{ F, \bar Q_2\} = -Q_2$,
and $H= \{ Q_1, \bar Q_2\}$.
The regularized nonchiral partition function,
generalizing \qllnrmi,  can be
defined as:
\eqn\sqmi{
\log \CZ_{\rm nonchiral}
 = \int_{\Lambda^{-1}}^\infty {dt \over  t} \Tr (-1)^F F e^{-t H}
}
Subtracting the divergences for $\Lambda^{-1} \rightarrow 0$
introduces the standard ambiguity by local counterterms.

Using standard manipulations and the
properties
$Q_1^2 = \bar Q_2^2 =0$ we find that
under a variation $\bar Q_2 \rightarrow \bar Q_2 +
 [ \bar Q_2, \epsilon] $ holding $Q_1$ fixed,
we have:
\eqn\vareffone{
\dzo_\epsilon  \log \CZ_{\rm nonchiral}
=
\dzo_\epsilon
\int_{\Lambda^{-1}}^\infty {dt \over  t} \Tr (-1)^F F e^{-t H}
= \lim_{\Lambda \rightarrow \infty}
\Tr (-1)^F \epsilon e^{-  H/\Lambda}
}
Let us give two examples of results easily
obtained in this way.

\subsubsec{Gauge cocycle}

First, let us put $\bar \mu=0$,
  work in $2n$ dimensions with a Euclidean
metric, and consider
``chiral gauge transformations''
$(a, \bar a) \rightarrow (a, \bar a + \pab \epsilon)$.
We then recognize \vareffone\ as the formula used
in the evaluation of anomalies using the
``Fujikawa method, '' and hence the answer
for the infinitesimal cocycle
follows immediately: \foot{ Note that this appears
to be a consistent, not a covariant
anomaly. However, under holomorphic
gauge transformations acting only on
$\bar a$ it is in fact a consistent anomaly.}
 \eqn\infgauge{
{ 1 \over  (2\pi)^n n!}  \int_{X_{2n}} \Tr \epsilon \bigl(
F(a, \bar a)\bigr)^n
}
As in the previous two sections we observe a
violation of $Q$-symmetry since the infinitesimal
cocycle depends on $a$, and as in the previous
two sections we can add a local counterterm to
remedy this.
\foot{The explicit form of this counterterm,
$\beta(a,\bar a) $ is given in the next
section.} The resulting cocycle is simply
obtained by putting $a=0$ to give:
\eqn\infgaugei{
C(\epsilon,\bar a)=
{d \over  d \xi} \biggl\vert_{\xi=0}
\Biggl[
(2\pi i) \int_X \ch (
d\bar z^{\bar a} dz^c F_{\bar a c} +
\xi \epsilon)\Biggr]
=
{ 1 \over  (2\pi i)^n n!}  \int_{X_{2n}} \Tr \epsilon \bigl(\p
\bar a)^n
}

\subsubsec{Diffeomorphism cocycle}

We now  indicate how one may
include the $\bar \mu$ dependence in our
results.
For simplicity we take the simplest
case of $\bar a_1 = \bar a_2 = \bar\mu_2 = 0$,
but work in $2n$ dimensions. In our supersymmetric
quantum mechanics the
  bosonic coordinates are $\phi^i, \bar \phi^{\ib} $
with fermionic partners $\psi^i , \bar \psi^{\ib}$ and
Q-symmetries:
\eqn\sqmii{
\eqalign{
Q_1: \bar \phi^{\ib}   \rightarrow \bar \psi^{\ib}\qquad
&
\qquad
\bar \psi^{\ib}  \rightarrow  0 \cr
  \phi^{i}   \rightarrow  \bar\psi^{\jb} \bar\mu_{\jb}^{~i} \qquad
& \qquad
  \psi^{i}   \rightarrow  \dot {\phi}^i - \dot {\bar \phi}^{\jb}
  \bar\mu_{\jb}^{~i} - \psi^k \bar \psi^\jb\p_k  \bar\mu_{\jb}^{~i}
 \cr
\bar Q_2:   \phi^{i}   \rightarrow   \psi^{i} \qquad & \qquad
\bar \phi^{\ib}   \rightarrow 0\cr
  \psi^{i}   \rightarrow 0 \qquad & \qquad
\bar \psi^{\ib}  \rightarrow \dot {\bar \phi}^{\ib} \cr}
}
One can easily check that $Q_1^2=\bar Q_2^2=0$ using
the Kodaira-Spencer equation. Moreover,
$\{Q_1, \bar Q_2\} $ acts on the fields as
${d \over  dt}$ and the action is
\eqn\sqmiii{
S_{SQM} = \{ \bar Q_2, \{ Q_1,\int  \dot \phi^i \p_i K \} \}
}
where $K$ is the Kahler potential.

The result for \vareffone\ is easily obtained
following closely the manipulations in
section 11 of \agwitt. The result is
simply expressed in terms of the
 Todd class with the curvature
(in holomorphic frame indices $a,b,...$) shifted by:
 \eqn\replace{
R^d_{~~b \bar a c}   \rightarrow
 R^d_{~~b \bar a c}
+ \p_b\p_c \bar \mu_{\bar a}^{~~d}
 }
Explicitly, taking the metric to be flat:
\eqn\diffcyc{
C(v,\bar \mu)=
{d \over  d \xi} \biggl\vert_{\xi=0}
\Biggl[
\int_X Td (
d\bar z^{\bar a} dz^c \p_b\p_c \bar \mu_{\bar a}^{~~d} +
\xi \p_b v^d)\Biggr]
}

\subsubsec{Integrated anomaly}

Combining \vareffone\ and \infgauge\ we obtain
the nonchiral partition function
$\CZ_{\rm nonchiral}[\bar a, a]$ by integrating
both variations. Putting $a=0$ we can then obtain
the chiral partition function. The resulting action
is described in section seven below.
Similar remarks apply to the diffeomorphism case.

\newsec{Holomorphic descent and Bott-Chern classes}

In the previous section we have argued that
the chiral $bc$ partition function is simply
expressed in terms of a group cocycle
for the complexified gauge group taking
values in functionals of differential operators
$\pb_{\bar a}$ squaring to zero.
The proper mathematical objects for understanding
this are Bott-Chern holomorphic
secondary characteristic classes
\ref\bottchrn{
R. Bott and S.S. Chern, ``Hermitian vector
bundles and the equidistribution of the
zeroes of their holomorphic sections,''
Act. Math. {\bf 114}(1965)71}\doniii\donii\BGS.

Let us begin by recalling
 the usual construction of secondary
characteristic classes. If $P(F)$ is an
invariant polynomial on  the Lie algebra $\lieg$
of a gauge group  then, under an
arbitrary variation of the gauge field:
\eqn\botti{
\delta P(F) = d \bigl[ P'(\delta A, F) \bigr]
}
Integrating this along a path between
two connections gives the Chern-Simons
form:
\eqn\chrnsm{
d CS[A_1; A_0] = P(F_1) - P(F_0)
}
{}From gauge variation of $A_1$, holding
$A_0$ fixed we obtain the standard descent
tower:
\eqn\realdesc{
\eqalign{
\delta_\epsilon CS[A_1;A_0] & = d \omega^1_{2n}[\epsilon, A_1, F_1]\cr
\cdots & \cdots \cr} }
leading to the standard group cocycle for gauge
anomalies.

\subsec{Bott-Chern classes}

Let us now assume that $X$ has   a
complex structure and that $F$ is of type
$(1,1)$: that is
$(a,\bar a) \in \CA^{1,1}$,   may be written as:
\eqn\untryfrm{
\eqalign{
a &= g_L^{-1} \p g_L \cr
\bar a & = - (\pb g_R) g_R^{-1}\cr}
}
For a unitary connection $g_R= g_L^\dagger$,
but we will not assume this in general.
This connection defines a holomorphic vector
bundle $\CE$ with holomorphic framing
defined by $\pab {\bf \vec e}=0$. It follows from
\untryfrm\ that
$F =g_R \pb(h^{-1}\p h) g_R^{-1} =
-g_L^{-1} \p(\pb h h^{-1}) g_L$
where $h=g_L g_R$. For unitary
connections $h$ is a hermitian metric
on the holomorphic bundle $\CE$.
Thus we may equivalently think of the
background data as a connection in
$\CA^{1,1}$  or as a quadratic form on
a holomorphic bundle $\CE$. Unitary
connections correspond to positive
hermitian forms.

Using the complex structure we may decompose
\eqn\bottip{
\delta P(F) = \pb \bigl[ P'(\delta a^{1,0} , F) \bigr]
+
\p \bigl[ P'(\delta \bar a^{0,1} , F) \bigr]
}
under an arbitrary variation of connection.
Moreover, if we are taking a
 variation of $A$ within $\CA^{1,1}$ then
the variation
is formally equivalent to a gauge transformation:
\eqn\chiralvars{
\eqalign{
\dzo_{\epsilon_L} a = \p_a \epsilon_L \qquad &
\qquad \dzo_{\epsilon_L} \bar a = 0 \cr
\doz_{\epsilon_R} a = 0  \qquad &
\qquad \doz_{\epsilon_R} \bar a = -\pb_{\bar a} \epsilon_R \cr}
}
and hence, by the descent formalism a variation
$\delta P(F)$ along $\CA^{1,1}$
is in fact $\p \pb$ exact:
\eqn\bottii{
\delta P(F) =( \delta^{1,0} + \delta^{0,1})
 P(F)= \pbar \p  \delta \CR(h)
}
Integrating this equation along a path
gives the
Bott-Chern
holomorphic secondary classes:
\eqn\btchrn{
\pb\p \CR(h_1; h_0) = P(F(h_1)) - P(F(h_0))
}
which should be viewed as the analog of
\chrnsm. The functional $\exp\int_{X_{2n}}\CR(h_1; h_0)$ is
the ratio of two nonchiral partition functions
of the type described in the previous section.

\subsec{Holomorphic descent }

It is important to stress that the
 Bott-Chern classes are {\it not} precisely
the group cocycles that we want. These
lead to the infinitesimal gauge cocycles
\eqn\gauge{
\int \Tr \epsilon F(a,\bar a)^n
}
Recall that we must maintain $Q$-symmetry, and
consequently  $a$-independence.
This is easily remedied
 by adding the coboundary $\beta$
mentioned in the previous section.
Indeed, let $(a,\bar a) $ be on $\CA^{1,1}$.
Then
\eqn\splteff{
\Tr(F^n) = \Tr(\p \bar a)^n + \Tr (\pb a)^n
+   \pb\p \beta(a, \bar a)
}
where $\beta$ is local and given explicitly
by:
\eqn\explcbet{
\eqalign{
\beta(a, \bar a)   =
\int_0^1 dt \int_0^1 ds \sum_{j=0}^{[(n-2)/2] }
& \qquad\qquad\qquad\qquad\qquad \cr
{(t^2-t)^j (s^2-s)^j \over  (j!)^2}
&
Str \bigl[ (a^2)^j , (\bar a^2)^j, a, \bar a,
(F_{s,t})^{n-2j-2} \bigr] \cr
F_{s,t} & = t \pb a + s \p \bar a + st [ a, \bar a] \cr}
}
where $Str$ is the graded symmetrized trace.
This can be more elegantly written as:
\eqn\elegant{
\int_0^\infty d\tau e^{-\tau} \int_{X_{2n}\times I \times I}
\Tr \exp[ \CF^{0,2} + \CF^{2,0} + \tau \CF^{1,1} ]
}
where $A(x,s,t) = t a(x) + s \bar a(x) $ and $ds,dt$ have
type $(1,0)$ and $(0,1)$ respectively.

The addition of $\doz \beta$ (where $\doz$
takes a gauge variation of $\bar a$ holding
$a$ fixed)  converts \gauge\ to the holomorphic
cocycle
\eqn\holococ{
\int \Tr \epsilon(\p \bar a)^n
}
The condition $F^{0,2}(\bar a) = 0$ is crucial
here.

The infinitesimal cocycle \holococ\
  can also be derived directly
by the following
 holomorphic version of the descent procedure.
Again let $P$ be an invariant polynomial on the
Lie algebra $\lieg$. We consider the
$(n+1,n+1)$-form $P(\p \bar a)$. We assume
$F^{0,2}=0$ so  it is $\p$ and $\pb$ closed.
Under an infinitesimal gauge transformation
\eqn\ddbdsc{
\eqalign{
P(\p \bar a) & \rightarrow \pb \p \wp(\epsilon, \p \bar a)\cr}
}
The $(n,n)$ form
$\wp(\epsilon, \p \bar a)$ is a Lie algebra cocycle.
Equivalently, we may write:
\eqn\holdesc{
P(\p \bar a) = \p P_1(\bar a, \p \bar a)
}
and, under gauge transformation
$\bar a \rightarrow \bar a + \pb_{\bar a} \epsilon$:
\eqn\holdescii{
P_1 \rightarrow P_1 +
\pb[\wp(\epsilon, \p \bar a)] + \p \xi^{n-1,n+1}
}
where $\xi^{n-1,n+1}$ is some form.
In this way we recover the holomorphic
cocycle $\wp$.

\newsec{Chiral Cocycle Theories }

In this section we briefly examine the
effective actions induced by $bc$ systems. These
generalize the $WZW$ theories of
two and four dimensions.

\subsec{Gauge transformations}

{}From the infinitesimal cocycle of the
previous section we obtain
the equation of motion
\foot{While this equation is rather peculiar it is worth noting
that it is not more than second derivatives in any
one coordinate and hence probably has a reasonable
initial value problem. }
\eqn\highdim{
(\pb (g^{-1} \p g))^n = 0
}
The corresponding group cocycle
is easily written as follows.
  Let us extend
the field $g$ from $X_{2n}$ to $X_{2n}\times D$
where $D$ is a disk and $g=1$ on the outer rim of
the disk. We may then simply take:
\eqn\gencocy{
\Gamma[g] =
\int_{X_{2n} \times D}
 {dw \over  w} \wedge \Tr \bar \ell (\p \bar \ell)^n
}
where $\bar \ell = g^{-1}\pb g$. The
action is independent of small changes of
extension and a variation of $g$ leads to
the equations of motion \highdim. Moreover,
$\Gamma[g]$ differs from $\int_{X_{2n}\times I}  \Tr (g^{-1} dg)^{2n+1} $
by a local functional of $g$ and hence
$\exp \kappa \Gamma[g]$ is single-valued for an appropriate
constant $\kappa$.

One can write the action \gencocy\ in
many ways. One way to make contact with
the standard WZ term is to take an extension
$g(x,w)$ which only depends on the radial
coordinates: $g(x,w) = g(x, \vert w \vert)$. The
integral over the disk reduces to an integral
over an interval. This presentation also makes
contact with higher-dimensional
Chern-Simons theories
\fs\ref\fnrs{V. Fock, N. Nekrasov,
A. Rosly, and K. Selivanov, ``What we
think about higher-dimensional Chern-Simons
theories,'' Published in Sakharov Conf.1991:465-472}.
In four dimensions
\eqn\rsngri{
\Gamma[g]  = \int_{X_4} \Tr\Biggl[
2(\ell \bar \ell - \bar \ell   \ell ) \pb \ell + (\bar \ell   \ell)^2 \Biggr]
+ {2 \over  5} \int_{X_4\times I} \Tr (g^{-1} dg)^5
}
where $\ell=g^{-1}\p g, \bar \ell = g^{-1}\pb g$.
In the abelian case the chiral
cocycle Lagrangian simplifies to
\eqn\plbsky{
4 \int_{X_4} \p \phi \pb \phi \p\pb \phi
}
for $g=\exp \phi$. This is closely related to the
Plebanski action for self-dual gravity \ogvf.
\foot{Quantization of this action and of similar
actions has been studied by A. Gerasimov.}

These actions satisfy some curious properties.
For example, a four-dimensional extension of
the Polyakov-Wiegmann formula is
\eqn\fdpw{
\eqalign{
\Gamma[g_1 g_2] & = \Gamma[g_1] + \Gamma[ g_2] \cr
& -2 \int_{X_4} \Tr\Biggl[
  \ell_1(\bar r_2 \p \bar r_2 + \p \bar r_2 \bar r_2) +
\bar r_2 (   \ell_1 \pb   \ell_1 + \pb   \ell_1   \ell_1)
-3 \bar r_2   \ell_1 \bar r_2   \ell_1 \Biggr]
\cr}
}
where $r= \p g g^{-1}, \bar r = \pb g g^{-1}$.
This equation might be of use in trying to define
a quantum version of the theory.

\subsec{Nontrivial background metric}

It is instructive to compute a nonchiral partition
function $\CZ_{\rm nonchiral}[a,\bar a ,\gamma]$ in
the presence of a nontrivial background
\k\ metric $\gamma$. This metric enters the
partition function through the measure.
(Here we work at fixed complex structure.)
As explained in sections five and six, the
variation of $\CZ_{\rm nonchiral}[a,\bar a ,\gamma]$
can be extracted from holomorphic descent.
Therefore, in four dimensions we start with
the differential form following from the
characteristic class:
\eqn\dolbe{
\eqalign{
 Td(T^{1,0}X_4) ch(E)\biggl \vert_6 & =
{1\over  3!} \Tr \bigl({i  F\over 2\pi  } \bigr)^3
+
{1 \over  2!} \Tr \bigl({i F\over  2\pi  } \bigr)^2 \half c_1(R) \cr
& +
\Tr \bigl({i F\over  2\pi  } \bigr) \bigl[{1 \over  8} \bigl(c_1(R) \bigr)^2
- {1 \over  24} \Tr \bigl({i R\over  2\pi  } \bigr)^2\bigr]\cr
&+
{r \over  48} c_1(R)
\bigl[ c_1(R)^2 - \Tr \bigl({i  R\over 2\pi  } \bigr)^2
\bigr] \cr}
}
Here the curvature on $T^{1,0}X$ is
  $R=\pb(\gamma^{-1} \p \gamma)$ and we write
$c_1(R)   = {i  \over  2 \pi  } \pb \p \log \det \gamma
\equiv {i  \over  2 \pi  } \pb \p \sigma = {\Tr} \bigl({iR\over{2\pi}}\bigr)$.
The trace of powers of $F$ is taken in the representation of $G$,
which corresponds to the choice of $E$. In the subsequent formulae
this is implicitly understood.

\subsubsec{Nonchiral partition function}

We now use \dolbe\ to split holomorphically
the nonchiral partition function on $\CA^{1,1}$:
\eqn\nonchir{
\log \CZ_{\rm nonchiral}[a,\bar a ,\gamma] =
\Gamma[g_L,\gamma] + \bar \Gamma[g_R, \gamma]
+ \tilde \beta(a, \bar a, \gamma)
}
where $g_L,g_R$ are defined in \untryfrm.
The chiral splitting function now becomes
\eqn\newchspl{
\tilde \beta = {1 \over  2!} ( {i\over  2 \pi } )^2 \beta(a,\bar a)
+
 {1 \over  2!} ( {i\over  2 \pi } )^2 \int \Tr(a \bar a) c_1(R)
}
Hence we find:
\eqn\bckgnd{
\Gamma[g;\gamma] = \Gamma[g]  +  S_{c_1(R)} [g] +
\int_{X_4}
 \biggl[ {1 \over  4} (\Tr R )^2
- {1 \over  12} \Tr R^2 \biggr]
\log \det g
}
where for any two-form $\Omega$ we define:
\eqn\wzwiiv{
S_\Omega[g] \equiv
-{i \over  4 \pi} \int_{X_4}
\Omega\wedge  {\Tr} \bigl(g^{-1} \p g \wedge g^{-1} \pb g\bigr)
+ {i \over  12 \pi} \int_{X_5} \Omega\wedge {\Tr}(g^{-1} d g)^3,
}
and $\log \det g$ in \bckgnd\ is to be understood as $\Tr_{E} \log g$.
When $\Omega$ is the
Kahler form of a Kahler metric \wzwiiv\ is
simply the ``\wzwf\ action'' studied in
\avatar.

\subsubsec{On chiral partition functions}

Equation \bckgnd\ gives $\log\CZ_{ch}$
as a function of $\bar a$ in a background
\k\ metric $\gamma$.
The introduction of a nontrivial metric
brings up the rather subtle issue of when cocycles
should be regarded as trivial or not. A closely
related  and equally subtle point is the possibility
of a $Q$-anomaly in a background Kahler
field. By formal $Q$-symmetry the
second and third terms in \bckgnd\ should
be absent -- the dependence of $\CZ_{ch}$
on the Kahler metric is $Q$-exact.
Nevertheless, the cocycles associated with
these terms appear to be nontrivial.
In fact, however,
if  $c_1(R)^2 + c_2(R)$
can be written as
$\p \pb \alpha_1 $ (as is always true locally)
then we may integrate by parts in the
third term to obtain the local counterterm
\eqn\trivi{
\doz \int \alpha_1 \Tr \p
\bar a = \int \p\pb \alpha_1 \log \det g
}
By the same token, holomorphic factorization
cannot determine this term uniquely. There
is always the possibility of adding a term
$\int \p \pb \alpha \log \det g$ to the
action $\Gamma[g] $.

In an similar way, the second term
in \bckgnd\ may be regarded as following
from a ``trivial'' cocycle due to the identity:
\eqn\trivii{
\doz \int \p \alpha_2 \Tr(\bar a \p \bar a)
= 2 \int \pb\p \alpha_2 \Tr(\epsilon \p \bar a)
}
Clearly, one can add $\doz$-exact
terms, local  in $\bar a$ and the
external metric $\gamma$ to the chiral cocycle
Lagrangian. This point,
which plays a role in making contact with the
$\CN_{ws}=2$ string needs further clarification.

\subsubsec{Dependence of the nonchiral
partition function on the background K\"ahler metric}

If we wish to regard the K\"ahler
metric as dynamical then we must include
the contribution to the Bott-Chern class
coming from the last term in \dolbe.
These are not determined by holomorphic
splitting in $a,\bar a$, but are determined
from the family index theorem computations.
Thus, if we compare the
 {\it nonchiral} partition function in a background
K\"ahler metric $\gamma_{i \bar j} dz^i d\zb^{\bar j} $
to the partition function on a
flat background we find  \BGS:
\eqn\fdconfmly{
\Gamma^{\rm conf} [\gamma] =
   {1 \over  96} {1 \over  (2 \pi)^2}
\int\Biggl[  \p \sigma \pb \sigma (\pb \p \sigma)
+\pb \sigma
\Tr (\gamma^{-1} \p \gamma)(\pb (\gamma^{-1} \p \gamma)) \Biggr]
}
where $\sigma = \log\det \gamma$.
This is the higher-dimensional analog of the
conformal anomaly.
Again, the metric dependence comes from the measure of
the non-chiral system, as is confirmed by the locality of \fdconfmly.
The higher dimensional
analogue of the Polyakov action is
obtained by considering the
difference:
\eqn\polyak{
\Gamma^{\rm conf} [\gamma ]
- \Gamma^{\rm conf} [\hat \gamma ]
}
For example, $\int \p \sigma\pb \sigma\p\pb \sigma$ is
promoted to
\eqn\promoted{
{1\over{96}}\int \Biggl[
{1\over{ 4{\pi}^2}} \p \sigma\pb \sigma\p\pb \sigma
+
{3\over{2\pi i}} c_{1}(\hat R)  \p \sigma\pb \sigma
-
3 \bigl(c_{1}(\hat R) \bigr)^2 \sigma \Biggr]
}
where $\sigma = \log\det \gamma - \log \det \hat \gamma$.

\newsec{Diffeomorphism Cocycles}

As a step towards making the complex structure
dynamical we now state some results
showing how  considerations parallel to those
in the gauge theory  apply to
the group of chiral  diffeomorphisms.
Such diffeomorphisms are defined as follows.
Let $f$ stands for a vector of functions
$f^i(z_1, \dots , z_n, \bar z_1, \dots \bar z_n)$.
We consider the
 $z, \bar z$ to be independent
\foot{For example, we could work
in a real space of signature $(n,n)$. }
and consider the   group action:
$f\circ g(z,\bar z) \equiv f(g(z,\bar z), \bar z)$.
We will adopt the active viewpoint:
all functions whose arguments
are not explicitly indicated are
evaluated at $(z,\bar z)$.
We define
$J(f)_i^{~~j} \equiv \p_i f^j$ and
$\bar J(f)_{\bar i}^{~~j} \equiv \pb_{\bar i} f^j$.
 Dot $\cdot $ always denotes (finite dimensional) matrix
multiplication.

We define the Beltrami differential
$\bar \mu[f]$ to be the matrix:
\eqn\bdi{
\bar \mu[f]_{\bar i}^{~j} \equiv [\bar J(f) \cdot J(f)^{-1} ]_{\bar i}^{~j}
}
One easily verifies the composition law:
\eqn\mutrnm{
 \bar \mu[f_1\circ f_2] = \bar \mu[f_2] +
( \bar \mu[f_1] \circ f_2)\cdot J(f_2)^{-1}
}
and hence, under
  the infinitesimal right action
$f    \rightarrow f\circ \ell$ defined by
$\ell=z-v(z,\bar z)$ we have
 right action on the space of
Beltrami differentials:
\eqn\tmnmu{
\bar \mu[f_1] \rightarrow \bar \mu[f_1] - \pb v
- v^i \p_i \bar \mu[f_1]  + \bar \mu[f_1] \p v\quad .
}
Here $\p v= \p_i v^j$ is regarded as a matrix-valued
function.

Let us now try to find an action $\Gamma[f]$
whose variation under the right chiral diffeomorphism
is a local functional of the Beltrami differential.
Such an action must be the result for the
$\bar \mu$-dependence of the chiral partition
function \motivate.  The answer is easily motivated
by recalling the close relation between the
diffeomorphism and local Lorentz anomalies.

We define the action for diffeomorphisms
to be:
 \eqn\proposal{
\Gamma[f] \equiv \Gamma[J(F)^{-1} ]
}
where, to any $f$ we associate the inverse, denoted
as $F$ and defined by $f\circ F = F \circ f = z$
and
where on the right hand side we use the
action \rsngr\  for a $GL(n,\IC)$ matrix.
Under the right diffeomorphism action
$f \rightarrow f\circ \ell$, $\ell = z-v(z,\bar z)$, which is
equivalent to $\bar \mu \rightarrow \delta_v \bar \mu$
we have
\eqn\diffcob{
\delta_v \Gamma[f] = -12\int_{X_4} \Tr\biggl[
(\p v) (\CR_{\bar \mu})^2\biggr]
}
where  we define
a $(1,1)$-form $\CR_{\bar \mu}$ with values in
$gl(n,\IC)$ by
\eqn\mutoo{
\eqalign{
(\CR_{\bar \mu})_m^{~p} & \equiv dz^j d\zb^\kb
\bigl( \p_j \p_m \bar \mu_{\kb}^{~p} \bigr)\cr}
}
Comparing with \replace\ we see that we
have produced the group cocycle for
diffeomorphisms.

This action defines higher dimensional
generalizations of the
Virasoro algebra.
Indeed, since there are many invariant
polynomials on the Lie algebra
$gl(n,\IC)$ there are many generalizations
of the Virasoro algebra corresponding to
different representations of $gl(n,\IC)$
in \diffcob. For example, to get the
cocycle related to the Todd class in
4 dimensions one
must choose the virtual representation $R$
of $U(2)$
such that $\Tr_R(x) = { 1\over  48} [(\Tr_F (x))^3 - \Tr_F(x) \Tr_F(x^2)]$,
where $F$ is the fundamental of $U(2)$.
This concludes the computation of
the $\bar\mu$ dependence of the chiral partition
function for four-dimensions.

\newsec{Hypermultiplets and self-dual geometry}

In this section we restrict attention to four
dimensions.  For $bc$ in
a general representation of
the gauge group  the equation \highdim\
is fourth-order in derivatives.
However, there is one very
special class of representations leading
to more conventional field theories.
Let us choose a vector bundle of
the form $(E\oplus E^*) \otimes H$ where
$H$ is a  line bundle.
In more physical
terms, we consider a gauge group
$U(N) \times U(1)$ with left-handed
fermions in the representation
\eqn\nicerep{
(\rho,+1) \oplus (\rho^*, +1)
}
Given the relation of $bc$ systems
to  twisted
supersymmetry explained above
we see that the representation
\nicerep\  is
extremely natural yet special:
it is simply a hypermultiplet for
an $\CN=2$ $U(N)$
supersymmetric gauge
theory.

Note that the choice of
representation leads to a cancellation
of the standard nonabelian $(g^{-1} d g)^5$ term in
the group cocycle.
Denoting the $U(N) \times U(1)$
gauge field by $(A,A_{ab})$ we have, in
an $F^{0,2}=0$ background
$A^{0,1}=g^{-1} \pb g$,
$A_{ab}^{0,1}=\pb \varphi$.
The  action \rsngri\ evaluated
for the representation \nicerep\ is:
\eqn\answer{
\log \CZ_{ch}
= S_\Omega(g) + {1 \over  3!}{1 \over  (2 \pi i)^2}
\int \varphi (\p \pb \varphi)^2 +
\int \varphi {r \over  12} (c_1(R)^2 + c_2(R))
}
The first term is the \wzwf\ action defined in
\wzwiiv\ above and evaluated now for
 $\Omega={1 \over  2 \pi i} \p \pb \varphi$.
The equations of motion
(considering the metric to be nondynamical)
following from
\answer\ are:
 \eqn\frst{
\eqalign{
-{4 \pi^2 r  \over  12} (c_1(R)^2 + c_2(R)) +
\half \Tr (\pb (g^{-1} \p g))^2 + \half \p \pbar
\varphi \wedge \p \pbar \varphi  & = 0 \cr
(\p \pbar \varphi)\wedge \pbar (g^{-1} \p g) & = 0 \cr}
}

An immediate corrollary of this result is that
we have fermionized the algebraic sector
current correlators described in
\avatar, in analogy to fermion/boson
correspondence in \wzwt.
The connection between \nicerep\ and
the \wzwf\ theory was pointed out in
\avatar. It has also been discussed recently in
\ref\henneaux{Maximo Banados  ,  Luis J. Garay  ,  Marc Henneaux,
``The dynamical structure of higher dimensional Chern-Simons theory,''
hep-th/9605159}.

\subsec{Comparison with the $\CN_{ws}=2$ string}

The spacetime physics of the $\CN_{ws}=2$ string
has been related to self-dual geometry in
\marcus\ogvf.
The field content of the open $\CN_{ws}=2$ string
is given by a scalar field $K$  and adjoint-valued
fields $\pi^a(x)$. The equations of motion are
\marcus:
\eqn\opentwo{
\eqalign{
\p \pb K \wedge\biggl( \pb g^{-1} \p g \biggr) & = 0 \cr
\p \pb K \wedge \p \pb K & = \omega_0^2 + \ch_2(F) \cr}
}
Comparing with \frst\   we see that
identifying $K$ with $\varphi$ and $g$ with
$e^\pi$ we obtain similar equations. According to
the remarks in section 7.2.2 above we can introduce
local counterterms of the type \trivi\trivii\ to bring
\frst\ to precisely the form \opentwo\ of the open
  $\CN_{ws}=2$ string.

\newsec{Conclusions}

In conclusion we comment on a few
possible applications of the above
results.

The next step following the above investigation
is the formulation of the higher dimensional
analogs of the Knizhnik-Zamolodchikov
equations. This is currently under study.
Moreover, the relation to the \wzwf\ theory
opens the way to an investigation of
some representations of higher dimensional
current algebras. Holomorphic gauge transformations
lead to $Q$-exact changes in the action. We may thus
expect that the $Q$-cohomology of the
free field Hilbert space of a K\"ahler-twisted
chiral supermultiplet (more generally of a
$bc$ system on any complex manifold)
supports a representation
of higher loop gauge algebras. We hope to discuss this
elsewhere
\ref\susyrep{
A. Losev, G. Moore, N. Nekrasov, S. Shatashvili,
``Supersymmetry and the representations of
four-dimensional current groups,'' in progress.}.
It is quite likely that   representations related to
$bc$ systems will shed some light on the results of
Borcherds and Jorgenson and Todorov
concerning determinants of $ \pb$ operators
on Calabi-Yau manifolds \borchiv\jorgenson.

An important avenue for generalizing these
results is the generalization from
K\"ahler to Hermitian metrics. This can
probably be accomplished using the
supersymmetric quantum mechanics
techniques discussed above. This
generalization is necessary to discuss
the relation of the diffeomorphism anomalies
with the ``higher dimensional conformal anomalies''
such as \fdconfmly. This is currently under
investigation
\ref\morepromises{A. Losev, G. Moore, N. Nekrasov, S. Shatashvili, in
progress.}.

There are other interesting generalizations
defined,
for example adding superpotentials,
dynamical vectormultiplets, and supergravity.
It remains to be seen if these are tractable.
Some results in this direction have been
obtained in \johansen.

Also,  given the close analogy to
\pwf\plh,
it is natural to expect that
these results will have an application in
defining a four-dimensional analog of
bosonization.

Finally, these results might find some
application in understanding
string duality. First of all, the diffeomorphism
cocycles we have uncovered might be
of some use in quantization of $p$-branes
as quantum mechanical objects.
Second,
 the $\CN_{ws} = (2,1)$ heterotic string has recently
played a role in attempts to
achieve a deeper understanding of
string duality \kutmart. The
relevant geometry appears to be self-dual
geometry with torsion. It would be very
interesting to generalize the
results of this letter to include these
systems.

\centerline{\bf Acknowledgements}

We are grateful to
Anton Gerasimov for many
 comments related to the
issues discussed in this paper and
for sharing his insights with us. N.Nekrasov wishes
to thank the Erwin Schr\"odinger International Institute
for Mathematical Physics at Vienna where some of the
ideas of this paper have appeared and
especially K.Gawedzki and H.Grosse for hospitality.
He is also grateful to A.~ Rosly and V.~ Fock
for useful discussions. S. Shatashvili would like
to thank the CERN theory group for hospitality
during the completion of this paper.
The research of A.~ Losev was partially
supported by RFFI   grant   96-01-01101
and Volkswagen Stiftung. The research of G. Moore
is supported by DOE grant DE-FG02-92ER40704,
and by a Presidential Young Investigator Award. The
research of N. Nekrasov is supported by
the Porter Ogden  Jacobus fellowship.  The research  of S.
Shatashvili is supported
by DOE grant DE-FG02-92ER40704,  NSF CAREER award,  

DOE OJI award, and the
Alfred P. Sloane Foundation.

\listrefs
\bye